\documentclass[11pt,twocolumn]{article}

\usepackage{arxiv}
\usepackage{amsmath}
\usepackage{graphics}
\usepackage[utf8]{inputenc} 
\usepackage[T1]{fontenc}    
\usepackage{url}            
\usepackage{booktabs}       
\usepackage{amsfonts}       
\usepackage{nicefrac}       
\usepackage{microtype}      
\usepackage{lipsum}

\usepackage{amsfonts}
\usepackage{amssymb}
\usepackage{bbold}
\usepackage{booktabs}
\usepackage{ulem}
\usepackage{xr}
\usepackage[justification=centering ]{subcaption}
\usepackage[numbers]{natbib} 
\usepackage{graphicx}
\usepackage{amsmath}
\usepackage[usenames,dvipsnames]{xcolor}
\usepackage{multirow}
\usepackage{lscape}

\usepackage{amsfonts}
\usepackage{amssymb}
\usepackage{bbold}
\usepackage{booktabs}
\usepackage{ulem}
\usepackage{xr}
\usepackage[justification=centering ]{subcaption}
\usepackage{tikz}
\usetikzlibrary{shapes}
\usepackage{soul}

\newcommand{\Ecircle}{\raisebox{0.5pt}{\tikz{\node[draw,scale=0.5,circle,fill=white!100!yellow](){};}}}
\newcommand{\Etriangle}{\raisebox{0.5pt}{\tikz{\node[draw,scale=0.4,regular polygon, regular polygon sides=3,fill=white!100!yellow,rotate=0](){};}}}
\newcommand{\Esquare}{\raisebox{0.5pt}{\tikz{\node[draw,scale=0.5,regular polygon, regular polygon sides=4,fill=white!100!yellow](){};}}}

\newcommand{\Lcircle}{\raisebox{0.5pt}{\tikz{\node[draw,scale=0.5,circle,fill=white!60!yellow](){};}}}
\newcommand{\Ltriangle}{\raisebox{0.5pt}{\tikz{\node[draw,scale=0.4,regular polygon, regular polygon sides=3,fill=white!60!yellow,rotate=0](){};}}}
\newcommand{\Lsquare}{\raisebox{0.5pt}{\tikz{\node[draw,scale=0.5,regular polygon, regular polygon sides=4,fill=white!60!yellow](){};}}}

\newcommand{\LGcircle}{\raisebox{0.5pt}{\tikz{\node[draw,scale=0.5,circle,fill=white!50!brown](){};}}}
\newcommand{\LGtriangle}{\raisebox{0.5pt}{\tikz{\node[draw,scale=0.4,regular polygon, regular polygon sides=3,fill=white!50!brown,rotate=0](){};}}}
\newcommand{\LGsquare}{\raisebox{0.5pt}{\tikz{\node[draw,scale=0.5,regular polygon, regular polygon sides=4,fill=white!50!brown](){};}}}

\newcommand{\DGcircle}{\raisebox{0.5pt}{\tikz{\node[draw,scale=0.5,circle,fill=black!0!brown](){};}}}
\newcommand{\DGtriangle}{\raisebox{0.5pt}{\tikz{\node[draw,scale=0.4,regular polygon, regular polygon sides=3,fill=black!0!brown,rotate=0](){};}}}
\newcommand{\DGsquare}{\raisebox{0.5pt}{\tikz{\node[draw,scale=0.5,regular polygon, regular polygon sides=4,fill=black!0!brown](){};}}}

\newcommand{\Scircle}{\raisebox{0.5pt}{\tikz{\node[draw,scale=0.5,circle,fill=black!60!red](){};}}}
\newcommand{\Striangle}{\raisebox{0.5pt}{\tikz{\node[draw,scale=0.4,regular polygon, regular polygon sides=3,fill=black!60!red,rotate=0](){};}}}
\newcommand{\Ssquare}{\raisebox{0.5pt}{\tikz{\node[draw,scale=0.5,regular polygon, regular polygon sides=4,fill=black!60!red](){};}}}

\newcommand{\LBcircle}{\raisebox{0.5pt}{\tikz{\node[draw,scale=0.5,circle,fill=white!95!blue](){};}}}
\newcommand{\LBtriangle}{\raisebox{0.5pt}{\tikz{\node[draw,scale=0.4,regular polygon, regular polygon sides=3,fill=white!95!blue,rotate=0](){};}}}
\newcommand{\LBsquare}{\raisebox{0.5pt}{\tikz{\node[draw,scale=0.5,regular polygon, regular polygon sides=4,fill=white!95!blue](){};}}}

\newcommand{\Bcircle}{\raisebox{0.5pt}{\tikz{\node[draw,scale=0.5,circle,fill=white!45!blue](){};}}}
\newcommand{\Btriangle}{\raisebox{0.5pt}{\tikz{\node[draw,scale=0.4,regular polygon, regular polygon sides=3,fill=white!45!blue,rotate=0](){};}}}
\newcommand{\Bsquare}{\raisebox{0.5pt}{\tikz{\node[draw,scale=0.5,regular polygon, regular polygon sides=4,fill=white!45!blue](){};}}}

\newcommand{\DBcircle}{\raisebox{0.5pt}{\tikz{\node[draw,scale=0.5,circle,fill=white!5!blue](){};}}}
\newcommand{\DBtriangle}{\raisebox{0.5pt}{\tikz{\node[draw,scale=0.4,regular polygon, regular polygon sides=3,fill=white!5!blue,rotate=0](){};}}}
\newcommand{\DBsquare}{\raisebox{0.5pt}{\tikz{\node[draw,scale=0.5,regular polygon, regular polygon sides=4,fill=white!5!blue](){};}}}

\newcommand{\LGRcircle}{\raisebox{0.5pt}{\tikz{\node[draw,scale=0.5,circle,fill=white!95!green](){};}}}
\newcommand{\LGRtriangle}{\raisebox{0.5pt}{\tikz{\node[draw,scale=0.4,regular polygon, regular polygon sides=3,fill=white!95!green,rotate=0](){};}}}
\newcommand{\LGRsquare}{\raisebox{0.5pt}{\tikz{\node[draw,scale=0.5,regular polygon, regular polygon sides=4,fill=white!95!green](){};}}}

\newcommand{\GRcircle}{\raisebox{0.5pt}{\tikz{\node[draw,scale=0.5,circle,fill=white!45!green](){};}}}
\newcommand{\GRtriangle}{\raisebox{0.5pt}{\tikz{\node[draw,scale=0.4,regular polygon, regular polygon sides=3,fill=white!45!green,rotate=0](){};}}}
\newcommand{\GRsquare}{\raisebox{0.5pt}{\tikz{\node[draw,scale=0.5,regular polygon, regular polygon sides=4,fill=white!45!green](){};}}}

\newcommand{\DGRcircle}{\raisebox{0.5pt}{\tikz{\node[draw,scale=0.5,circle,fill=black!50!green](){};}}}
\newcommand{\DGRtriangle}{\raisebox{0.5pt}{\tikz{\node[draw,scale=0.4,regular polygon, regular polygon sides=3,fill=black!50!green,rotate=0](){};}}}
\newcommand{\DGRsquare}{\raisebox{0.5pt}{\tikz{\node[draw,scale=0.5,regular polygon, regular polygon sides=4,fill=black!50!green](){};}}}

\newcommand{\Ca}{\text{Ca}}

\newcommand{\Uw}{\text{U}_{\mbox{\scriptsize w}}}
\newcommand{\Bq}{\text{Bq}}

\newcommand{\muin}{\mu_{\mbox{\scriptsize in}}}
\newcommand{\muout}{\mu_{\mbox{\scriptsize out}}}
\newcommand{\mum}{\mu_{\mbox{\scriptsize m}}}

\newcommand{\effi}{f_i}

\newcommand{\Kal}{k_{\scriptsize \alpha}}

\newcommand{\kB}{k_{\mbox{\scriptsize B}}}
\newcommand{\kS}{k_{\mbox{\scriptsize S}}}

\newcommand{\WB}{W_{\mbox{\scriptsize B}}}
\newcommand{\WS}{W_{\mbox{\scriptsize S}}}

\newcommand{\cs}{c_{\mbox{\scriptsize s}}}
\newcommand{\mus}{\mu_{\mbox{\scriptsize s}}}
\newcommand{\mud}{\mu_{\mbox{\scriptsize d}}}

\newcommand{\tl}{t_{\mbox{\tiny L}}}
\newcommand{\tr}{t_{\mbox{\tiny R}}}

\newcommand{\ttilde}{\tilde{t}}
\newcommand{\tlcos}{t_{\mbox{\tiny L}}^{\mbox{\tiny cos}}}

\newcommand{\lst}{L_{\mbox{\tiny 1}}}
\newcommand{\lsh}{L_{\mbox{\tiny 2}}}
\newcommand{\dA}{d_{\mbox{\scriptsize A}}}
\newcommand{\dT}{d_{\mbox{\scriptsize T}}}

\newcommand{\Lx}{L_{\mbox {\tiny x}}}
\newcommand{\Ly}{L_{\mbox {\tiny y}}}
\newcommand{\Lz}{L_{\mbox {\tiny z}}}

\newcommand{\gammadot}{\dot{\gamma}}

\newcommand{\deltal}{\delta_{\mbox{\tiny L}}}
\newcommand{\deltar}{\delta_{\mbox{\tiny R}}}

\newcommand{\dav}{D_{\mbox{\tiny av}}}
\newcommand{\froll}{\gammadot_{\mbox{\tiny FRMS}}}

\renewcommand{\vec}[1]{\ensuremath{\boldsymbol{#1}}}
\renewcommand{\vec}[1]{\ensuremath{\mathbf{#1}}}

\title{\huge Loading and relaxation dynamics\\ of a red blood cell~\dag}

\author{
  Fabio Guglietta\\ 
  Department of Physics \& INFN, {\it University of Rome ``Tor Vergata''}\thanks{Via della Ricerca Scientifica 1, 00133, Rome, Italy}\\ 
Chair for Computational Analysis of Technical Systems (CATS), {\it RWTH Aachen University}\thanks{52056 Aachen, Germany}\\
Computation-based  Science  and  Technology  Research  Center, {\it The  Cyprus  Institute}\thanks{20 Konstantinou Kavafi Str., 2121 Nicosia, Cyprus}\\
  \texttt{fabio.guglietta@roma2.infn.it} \\
   \And
 Marek Behr\\
 Chair for Computational Analysis of Technical Systems (CATS), {\it RWTH Aachen University}$^{\dagger}$\\
  \And
 Giacomo Falcucci\\
 Department of Enterprise Engineering “Mario Lucertini,” {\it University of Rome ``Tor Vergata"}\thanks{Via del Politecnico 1, 00133, Rome, Italy}\\ 
 \AND
 Mauro Sbragaglia\\
  Department of Physics \& INFN, {\it University of Rome ``Tor Vergata''}$^{*}$\\ 
}

\begin{document}
\twocolumn[
\begin{@twocolumnfalse}
\maketitle
\begin{abstract}We use mesoscale numerical simulations to investigate the unsteady dynamics of a single red blood cell (RBC) subjected to an external mechanical load. We carry out a detailed comparison between the {\it loading} (L) dynamics, following the imposition of the mechanical load on the RBC at rest, and the {\it relaxation} (R) dynamics, allowing the RBC to relax to its original shape after the sudden arrest of the mechanical load. Such a comparison is carried out by analyzing the characteristic times of the two corresponding dynamics, i.e., $t_L$ and $t_R$. When the intensity of the mechanical load is small enough, the two kinds of dynamics are {\it symmetrical} ($t_L \approx t_R$) and independent of the typology of mechanical load (intrinsic dynamics); otherwise, in marked contrast, an {\it asymmetry} is found, wherein the loading dynamics is typically faster than the relaxation one. This asymmetry manifests itself with non-universal characteristics, e.g., dependency on the applied load and/or on the viscoelastic properties of the RBC membrane. To deepen such a non-universal behaviour, we consider the viscosity of the erythrocyte membrane as a variable parameter and focus on three different typologies of mechanical load (mechanical stretching, shear flow, elongational flow): this allows to clarify how non-universality builds up in terms of the deformation and rotational contributions induced by the mechanical load on the membrane. Finally, we also investigate the effect of the elastic shear modulus on the characteristic times $t_L$ and $t_R$. Our results provide crucial and quantitative information on the unsteady dynamics of RBC and its membrane response to the imposition/cessation of external mechanical loads.
 \vspace{1 cm}
\end{abstract}
\end{@twocolumnfalse}
]

\footnotetext{\dag~Electronic Supplementary Information (ESI) available: one PDF containing further details regarding the simulations performed; four videos showing the simulations performed. See DOI: 10.1039/D1SM00246E}



\section{Introduction}\label{sec:intro}
\begin{figure*}
\centering
\begin{tabular}{c c}
\includegraphics[width=.4\linewidth]{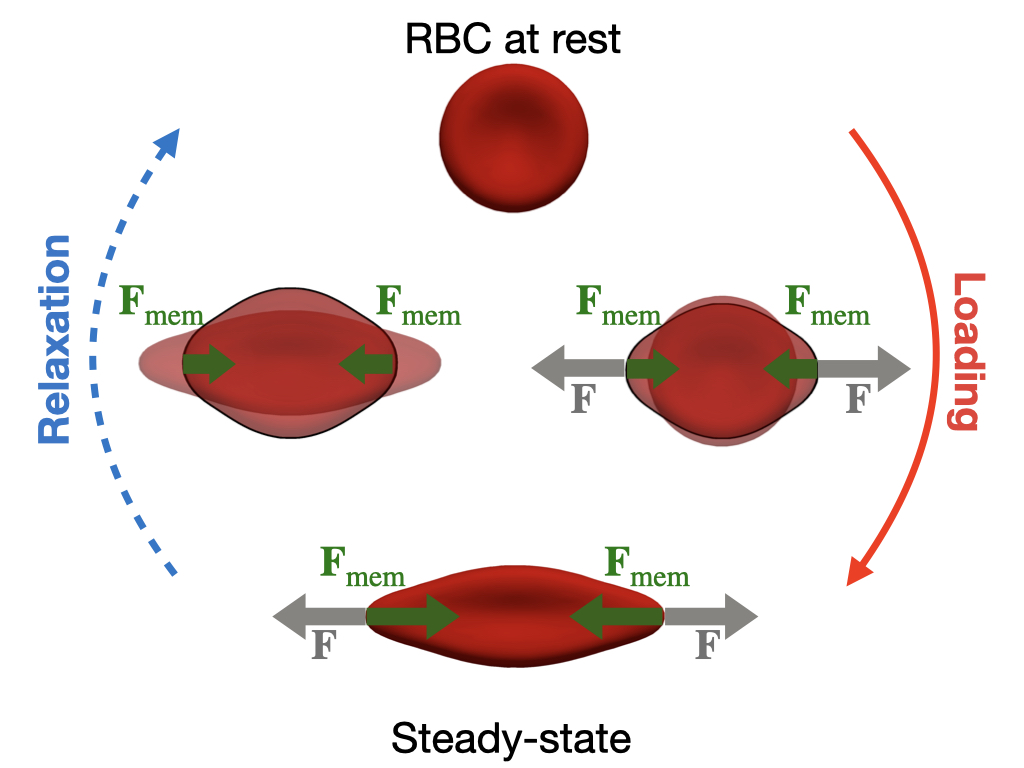} & \includegraphics[width=.4\linewidth]{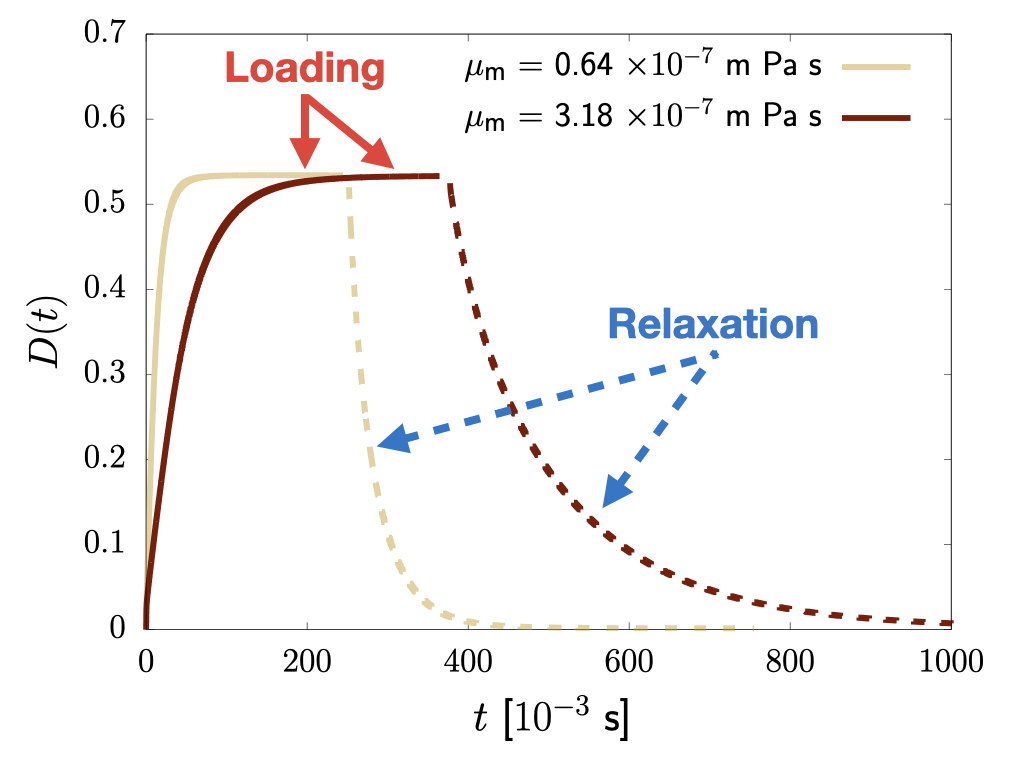}\\
\small (a) {\bf ST}retching {\bf S}imulation (STS) & \small (b) Stretching simulation (STS): deformation. \vspace{.5 cm}
\\
\includegraphics[width=.4\linewidth]{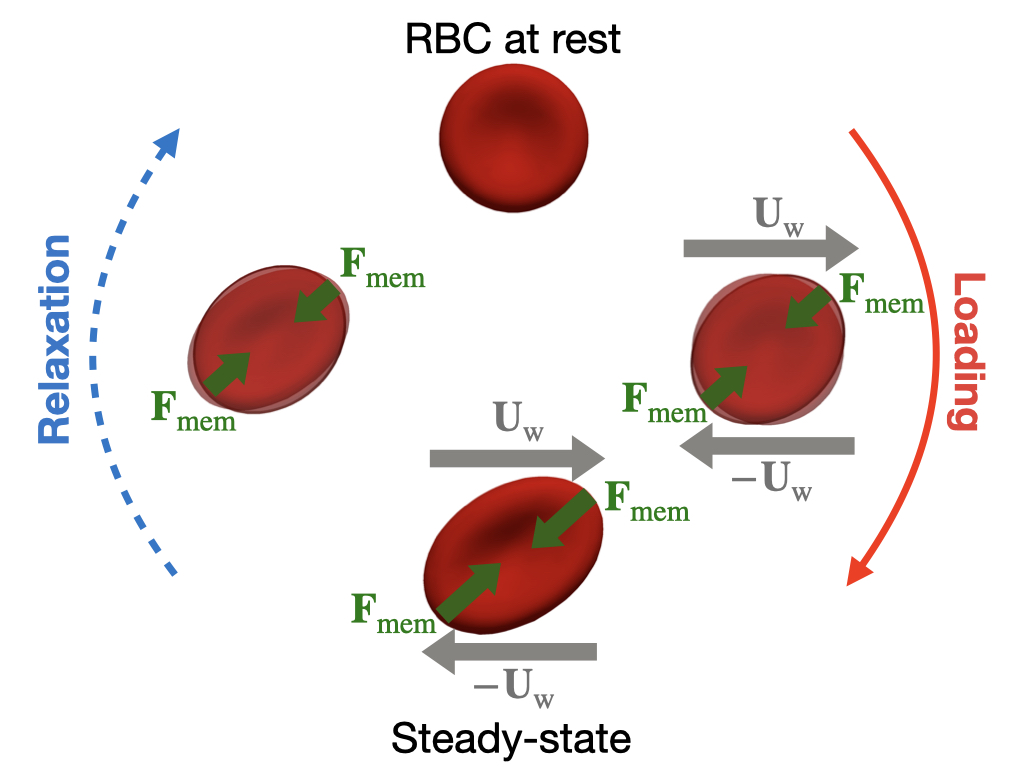} & \includegraphics[width=.4\linewidth]{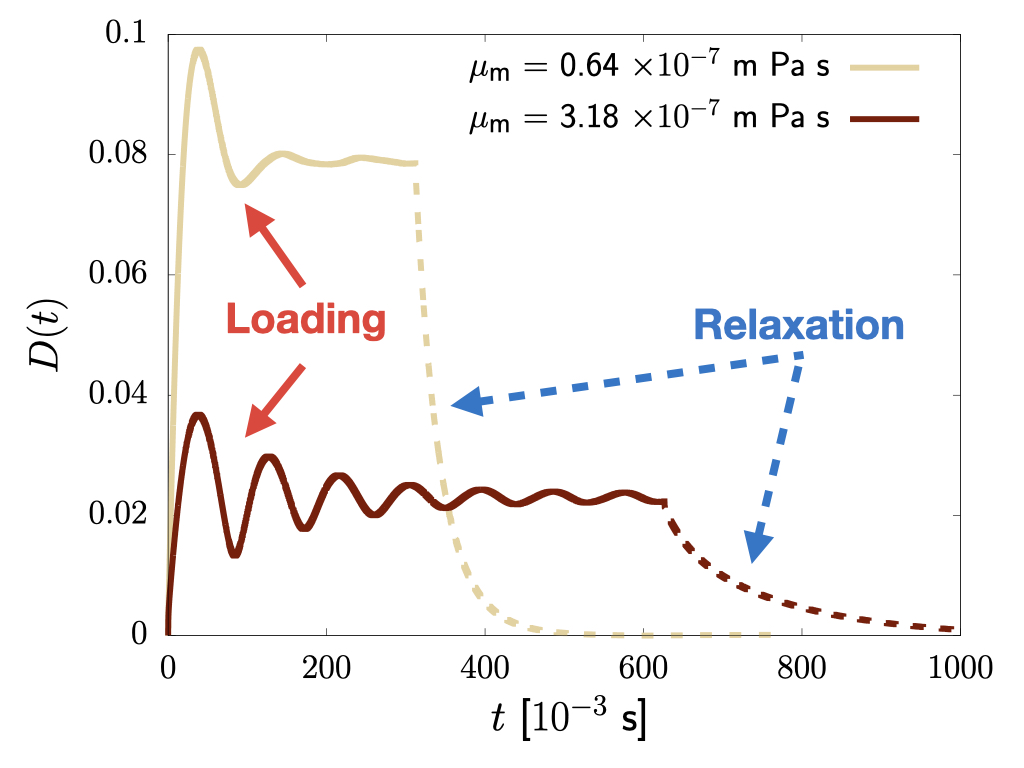}\\
\small (c) {\bf SH}ear {\bf S}imulation (SHS) & \small (d) Shear simulation (SHS): deformation. \vspace{.5 cm}\\
\includegraphics[width=.4\linewidth]{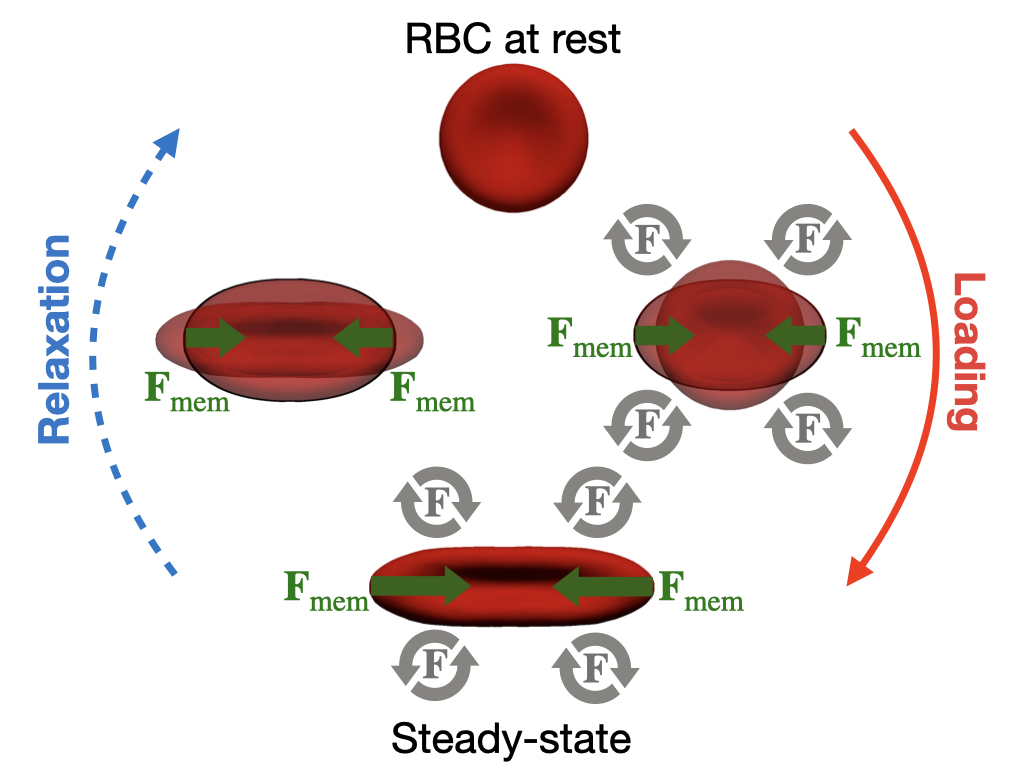} & \includegraphics[width=.4\linewidth]{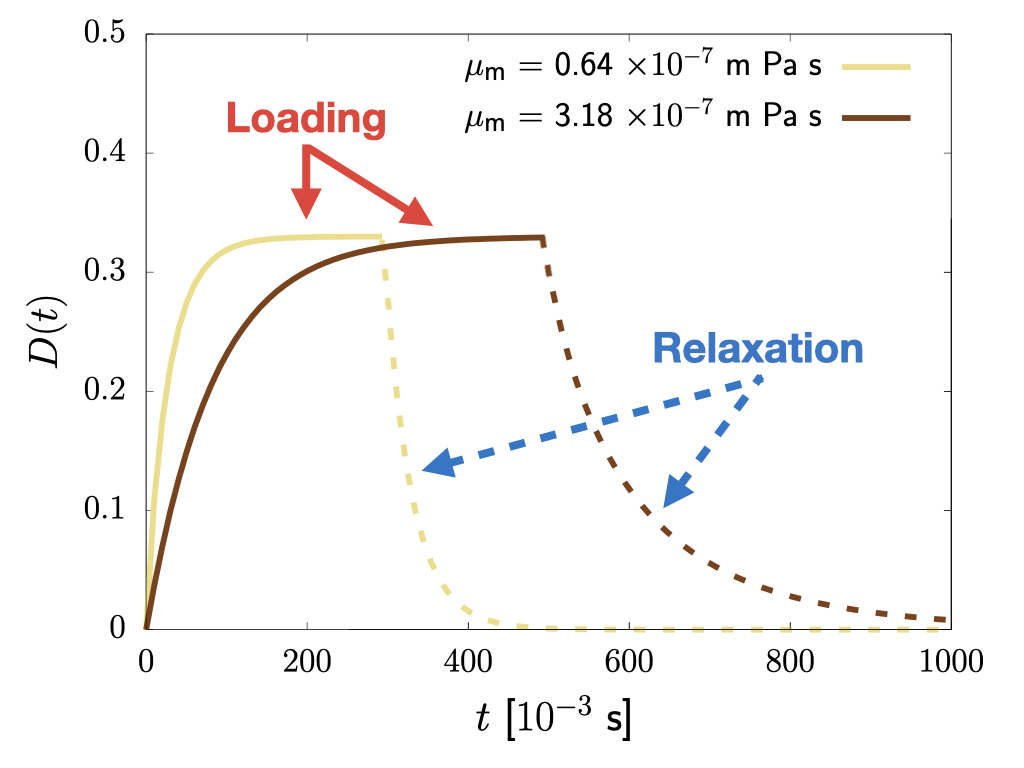}\\
\small (e) {\bf F}our-{\bf R}oll {\bf M}ill {\bf S}imulation (FRMS) & \small (f) Four-roll mill simulation (FRMS): deformation. \vspace{.5 cm}
\end{tabular}
\caption{Loading-relaxation (L-R) simulations for red blood cell (RBC) at changing the typology of mechanical load. \textit{\underline{Left panels}}: the three different L-R simulations investigated in the paper are sketched: grey arrows refer to the mechanical load, either an applied force $\vec{F}$ or an applied velocity $\Uw$, while the RBC membrane forces ($\vec{F}_{\mbox{\scriptsize mem}}$) are sketched with green arrows. In all simulations, the deformation $D(t)$, i.e., the ratio between the difference and the sum of the axial and transversal diameters (see Eq.~\eqref{eq:d}), is used to fit the loading and relaxation times ($\tl$ and $\tr$, respectively; see Eqs.~\eqref{eq:lst}~\eqref{eq:lsh}~\eqref{eq:r}). 
\textit{\underline{Right panels}}: we report the deformation $D(t)$ (see Eq.~\eqref{eq:d}) as a function of time for two values of membrane viscosity~$\mum$. 
{\it \underline{Panels (a-b)}}: we simulate the stretching with optical tweezers~\cite{art:suresh2005connections} (STS), in which two forces with the same intensity and opposite direction stretch the membrane in two areas at the ends of the RBC (see Sec.~\ref{sec:stretching}); the deformation $D(t)$ is reported for $F=90\times 10^{-12}$ N. 
{\it \underline{Panels (c-d)}}: deformation induced by simple shear flow (SHS), with $\Uw =  (\pm\dot{\gamma}/2,0,0)$, where $\gammadot$ is the shear rate (see Sec.~\ref{sec:shear}); the deformation $D(t)$ is reported for $\gammadot=86 \mbox{ s}^{-1}$. 
{\it \underline{Panels (e-f)}}: in the four-roll mill simulation (FRMS) we simulate four rotating cylinders to reproduce an elongational flow that deforms the membrane~\cite{malaspinas2010lattice} (see Sec.~\ref{sec:four-roll_mill}); the deformation $D(t)$ is reported for $\froll=80 \mbox{ s}^{-1}$.
Four videos showing these simulations are available (see ESI\dag).
\label{fig:sketch}}
\end{figure*}

Red blood cells (RBCs) are biological cells made of a viscoelastic membrane enclosing a viscous fluid (cytoplasm): their main features are the biconcave shape and the absence of a nucleus and most organelles, that allow them to carry oxygen even inside the smallest capillaries~\cite{popel2005microcirculation,skalak1969deformation,thesis:kruger}.
In fact, during circulation, RBCs deform multiple times, rearranging their shape to adapt to the physiological conditions of the blood flow.
The mechanical properties of RBC's membrane have been deeply investigated, both numerically~\cite{thesis:kruger,thesis:mountrakis,thesis:janoschek,gross2014rheology,art:kruger13,art:fedosov10,art:pan11,art:fedosov2010systematic,art:kruger14deformability,art:fedosov2011predicting,art:guckenberger2018numerical,fedosov2014multiscale} and experimentally~\cite{mills2004nonlinear,art:suresh2005connections,braunmuller2012hydrodynamic,chien1978theoretical,art:prado15,art:henon99,art:hochmuth79,art:baskurt96,art:bronkhorst95}. 
Research interest on the mechanical response of RBC's membrane was prompted by several reasons: among the others, the link between its properties and the erythrocyte's health conditions~\cite{art:kruger14deformability,art:suresh2005connections}, or the role played by the membrane dynamics in the design of biomedical devices\cite{murakami1979nonpulsatile,nonaka2001development,art:behbahani09,art:arora2006hemolysis}. 
A huge effort has been devoted to the characterisation of the time-independent properties of the membrane, for the study of the corresponding steady-state configurations.
In recent years, also the dynamical behaviour of RBCs has been investigated in several works, both numerically~\cite{zhu2019response,cordasco2017shape,cordasco2016dynamics,cordasco2014intermittency,art:fedosov10,fedosov2010multiscale,D0SM00587H,guglietta2020lattice} and experimentally~\cite{braunmuller2012hydrodynamic,art:prado15,art:henon99,dao2003mechanics}. When dealing with time-dependent properties of biological cells (or capsules, in general), the membrane viscosity plays a crucial role~\cite{li2020finite,li2020similar,art:yazdanibagchi13,D0SM00587H,guglietta2020lattice,graessel2021rayleigh,noguchi2005dynamics,noguchi2007swinging,barthes2016motion,diaz2000transient}. Evans~\cite{evans19891} showed that the RBC relaxation time is affected by both the membrane viscosity and the dissipation in the adjacent aqueous phases (i.e., cytoplasm and external solution); neglecting the membrane viscosity, i.e., $\mum=0$, he predicted a relaxation time $\tr\approx 1\times 10^{-3}$~s (also confirmed by numerical simulations~\cite{D0SM00587H}), a remarkably lower value compared to other works in the literature~\cite{art:baskurt96,art:prado15,braunmuller2012hydrodynamic,hochmuth1979red,D0SM00587H}. Several works aimed at the precise estimation of the value of the membrane viscosity~\cite{evans1976membrane,chien1978theoretical,hochmuth1979red,tran1984determination,art:baskurt96,riquelme2000determination,art:tomaiuolo11,braunmuller2012hydrodynamic,art:prado15,fedosov2010multiscale}, finding  that $\mu_m$ roughly ranges between $10^{-7}\mbox{ m Pa s}$ and $10^{-6}\mbox{ m Pa s}$. 
Such variability may be ascribed to different factors, e.g., the different theoretical models used to infer $\mum$~\cite{art:prado15,evans1976membrane}, the different experimental apparatuses (such as micro-pipette aspiration~\cite{evans1976membrane,chien1978theoretical,hochmuth1979red,braunmuller2012hydrodynamic}, microchannel deformation~\cite{art:tomaiuolo11}, or other setups~\cite{tran1984determination,art:baskurt96,riquelme2000determination}), etc.
As a matter of fact, although $\mum$ is an essential parameter to quantitatively characterise the time dynamics of RBCs~\cite{guglietta2020lattice,li2020similar}, its precise value has not been accurately determined so far, which warrants a parametric investigation.  Moreover, some earlier studies~\cite{keller1982motion} proposed to use an increased apparent viscosity ratio to account for the energy dissipation due to the presence of a viscous membrane: even though this assumption provides a qualitative description of the effects of the membrane viscosity, it does not account for a quantitative characterisation, as shown by recent studies~\cite{matteoli2021impact,li2020similar,guglietta2020lattice,noguchi2007swinging,noguchi2005dynamics}.
Our previous work~\cite{D0SM00587H} aimed to investigate the effect of membrane viscosity $\mum$ on the relaxation dynamics of a single RBC, and we found that increasing the value of $\mum$, as well as increasing the intensity of the loading strength, leads to faster recovery dynamics. Moreover, we simulated two experimental setups, i.e., the stretching with optical tweezers and the deformation due to an imposed shear flow, and we found a dependency on the kind of mechanical load when the strengths of load are large enough.\\ 
The relaxation dynamics, however, gives only a partial characterisation of the time-dependent response of RBCs to external forces: therefore, the loading process should be considered as well, as already pointed out in earlier literature papers. Chien {\it et al.}~\cite{chien1978theoretical} experimentally studied both the loading and the relaxation dynamics of RBC membrane through micro-pipette aspiration, providing evidence that the two dynamics are not symmetrical in certain conditions; however, a systematic study involving different stress values and different typologies of mechanical loads was not performed. Diaz {\it et al.}~\cite{diaz2000transient} studied the dynamics of a pure elastic capsule with a hyperelastic membrane deformed by an elongational flow: they focused on both loading and relaxation, finding an asymmetry. However, their model did not take into account the membrane viscosity and the asymmetry was not studied for different typologies of mechanical loads. Thus, although previous literature points to two distinct dynamics for loading and relaxation~\cite{chien1978theoretical,diaz2000transient,barthes2016motion}, a comprehensive parametric study on the effects of $\mum$ for different typologies of mechanical loads and flow conditions (such as simple shear flow or elongational flow) has never been attempted, so far.
This paper aims at filling this gap with the help of mesoscale numerical simulations. Indeed, for this kind of characterisation, numerical simulations can be thought of as the appropriate tool of analysis~\cite{thesis:kruger,thesis:mountrakis,thesis:janoschek,gross2014rheology,art:kruger13,art:fedosov10,art:pan11,art:fedosov2010systematic,art:kruger14deformability,art:fedosov2011predicting,art:guckenberger2018numerical,fedosov2014multiscale}, due to the obvious experimental difficulties in carrying out such systematic investigation~\cite{braunmuller2012hydrodynamic,chien1978theoretical,art:prado15,art:hochmuth79,art:bronkhorst95}. We provide a quantitative characterisation of loading and relaxation dynamics exploring three typologies of mechanical loads. To do this, we built three different simulation setups: the stretching simulation (STS), which simulates the deformation with optical tweezers~\cite{art:suresh2005connections} (see Fig.~\ref{fig:sketch}, panel (a)); the shear simulation (SHS), i.e., the deformation in simple shear flow  (see Fig.~\ref{fig:sketch}, panel (c)); the four-roll mill simulation (FRMS), where the deformation is induced by an elongational axisymmetric flow made by the rotation of four cylinders (see Fig.~\ref{fig:sketch}, panel (e)). These three numerical setups are chosen to inspect the different roles that the membrane rotation and/or membrane deformation have in the time-dependent dynamics. This information is summarised in Tab.~\ref{tab:summary}. First, we systematically study the characteristic times of both the loading ($\tl$) and the relaxation ($\tr$) processes and their ratio $\ttilde=\tl/\tr$, as a function of the load strength and membrane viscosity $\mum$. For small strengths, the two characteristic times are essentially equal and set by the value of $\mum$; however, for strengths large enough, the loading dynamics is found to be faster than the relaxation dynamics, leading to  a non-universal behaviour while changing the typology of the mechanical load. Such non-universal contributions are further characterised in terms of the importance of rotation and deformation of the membrane, according to the different load mechanisms. Some useful parametrizations for both $\tr$ and $\tl$ as a function of the membrane viscosities are also provided.
Finally, since different pathologies that cause the reduction of RBC membrane elasticity are known~\cite{art:kruger14deformability, art:suresh2005connections,suresh2006mechanical,brandao2003optical,briole2021molecular,brandao2003elastic,fedosov2011quantifying,luo2013inertia,ye2013stretching,hosseini2012malaria}, we also study the dependency of both the loading ($\tl$) and the relaxation ($\tr$) times as well as their ratio $\ttilde$ on the elastic shear modulus $\kS$, for a fixed value of membrane viscosity.\\
The paper is organised in the following way: in Sec.~\ref{sec:method} we provide some details on the numerical method used to simulate both the fluid and the membrane of the RBC; in Sec.~\ref{sec:loading_relaxation} we analyse the three simulated loading mechanisms (the stretching simulation, Sec.~\ref{sec:stretching}; the shear simulation, Sec.~\ref{sec:shear}; the four-roll mill simulation, Sec.~\ref{sec:four-roll_mill}); a detailed discussion section with comparisons between the loading mechanisms is provided in Sec.~\ref{sec:comparison}; finally, conclusions are reported in Sec.~\ref{sec:conclusion}.\\

\begin{table}[]
\centering
\begin{tabular}{|c|c|c|}
 \hline
Load Type & Rotation & Direct Forcing\\
     \hline
STS  & NO & YES  
\\
SHS  & YES & NO   
\\
FRMS & NO  & NO  
\\   
\hline
\end{tabular}
\vspace{0.5cm}
\caption{Summary of the main characteristics of the three kinds of applied mechanical loads (see Fig.~\ref{fig:sketch}). For each load type, we specify if the rotation is induced on the membrane while loading and if the forcing is directly applied on the nodes of the mesh used to discretise the membrane (otherwise, the membrane is forced indirectly via hydrodynamic flow).}\label{tab:summary}
\end{table}
\section{Numerical method} \label{sec:method}
We perform three-dimensional numerical simulations in the framework of the Immersed Boundary -- Lattice Boltzmann method (IB--LBM)~\cite{book:kruger,succi2001lattice}. The methodology, as well as the membrane model, are the same already used and validated in \cite{D0SM00587H}: here we report an essential summary.\\
The equation of motion for a fluid with viscosity $\mu$ is given by the Navier-Stokes (NS) equations:
\begin{equation}\label{eq:NS}
\rho\left(\frac{\partial\vec{u}}{\partial t}+(\vec{u}\cdot\pmb{\nabla})\vec{u}\right)=-\pmb{\nabla} p+\mu\pmb{\nabla}^2\vec{u}+\vec{F}\; ,
\end{equation}
where $\rho$ and $\vec{u}$ are the density and the velocity of the fluid, respectively; $p$ is the isotropic pressure; $\vec{F}$ is an external body force density. If the fluid is incompressible (as in the present work) the condition $\pmb{\nabla}\cdot\vec{u}=0$ holds.\\
In the LBM, instead of directly solving the NS equations by integrating Eq.~\eqref{eq:NS}, the fluid is represented by the so-called populations $f_i(\vec{x},t)$, that stand for the density of fluid molecules moving with velocity $\vec{c}_i$ at position $\vec{x}$ and time $t$. The populations evolve according to the Lattice Boltzmann equation:
\begin{equation}\label{LBMEQ}
\begin{split}
\effi(\vec{x}+\vec{c}_i\Delta t, t+ \Delta t) -  \effi(\vec{x}, t) = \\
=-\frac{\Delta t}{\tau}\left(\effi(\vec{x}, t)  - \effi^{(\mbox{\tiny eq})}(\vec{x}, t)\right)& + \effi^{(F)}\; ,
\end{split}
\end{equation}
in which $\Delta t$ is the discrete time step, $\tau$ is the relaxation time, $\effi^{(F)}$ is the source term that takes into account the force density (it has been implemented according to the ``Guo'' scheme~\cite{PhysRevE.65.046308}), and $ \effi^{(\mbox{\tiny eq})}$ is the equilibrium distribution function (we refer back to~\cite{book:kruger,succi2001lattice} for the details). The fluid density $\rho$ and the velocity $\vec{u}$ are given by:
\begin{equation}
\begin{split}
\rho(\vec{x}, t) = \sum_{i} \effi(\vec{x}, t)\; , \\ \rho\vec{u}(\vec{x}, t) = \sum_{i} \vec{c}_i \effi(\vec{x}, t)\; .
\end{split}
\end{equation}
The link between NS and LB equations (Eq.~\eqref{eq:NS} and Eq.~\eqref{LBMEQ}, respectively) is given by the following relation:
\begin{equation}
\mu= \rho \cs^2\left(\tau-\frac{\Delta t}{2}\right)\; ,
\end{equation}
where $\cs=\Delta x/\Delta t\sqrt{3}$ is the speed of sound. In the following, we considered both the lattice spacing $\Delta x$ and the time interval $\Delta t$ equal to 1.
\\
We simulate two fluids: one outside the membrane (the plasma, with viscosity $\muout = 1.2\times 10^{-3}$ Pa s) and one inside it (the cytosol, with viscosity $\muin = 6 \times 10^{-3}$ Pa s). The viscosity ratio is given by
\begin{equation}\label{visc_ratio}
\lambda = \frac{\muin}{\muout}\; ,
\end{equation}
providing $\lambda = 5$. We implement the parallel Hoshen-Kopelman algorithm to recognise which lattice sites are inside or outside the membrane (see~\cite{art:frijters15} for details). \\ 
The RBC membrane is described as a 3D triangular mesh of $\approx~4000$ elements, whose shape at rest is given by~\cite{evans1972improved} 
\begin{equation}
\begin{split}
z(x,y) = \pm \sqrt{1-\frac{x^2+y^2}{r^2}}\cdot \\ \cdot\left(C_0+C_1\frac{x^2+y^2}{r^2}+ C_2\left(\frac{x^2+y^2}{r^2}\right)^2\right)\; ,
\end{split}
\end{equation}
with $C_0 = 0.81\times 10^{-6} \mbox{ m}$, $C_1 = 7.83\times 10^{-6} \mbox{ m}$ and $C_2 = -4.39\times 10^{-6} \mbox{ m}$; $r=3.91\times 10^{-6} \mbox{ m}$ is the large radius. \\
The membrane is characterised by a resistance to shear deformation, area dilation and bending; the viscoelastic behaviour is implemented, as well. The first two terms form the {strain energy} $\WS$ are described by Skalak model~\cite{art:skalaketal73}:
\begin{equation}\label{eq:skalak}
\WS = \sum_j A_j\left[ \frac{\kS}{12}\left(I_1^2+2I_1-2I_2\right) +  \frac{\Kal}{12} I_2^2\right]\; ,
\end{equation}
where $\kS = 5.3\times 10^{-6} \mbox{ N m}^{-1}$~\cite{art:suresh2005connections} and $\Kal = 50 \kS$~\cite{thesis:kruger} are the surface elastic shear modulus and the area dilation modulus, respectively; $I_1 = \lambda_1^2+\lambda_2^2-2$ and $I_2 = \lambda_1^2\lambda_2^2-1$ are the strain invariants for the $j$-th element, while $ \lambda_1$ and $ \lambda_2$ are the principal stretch ratios~\cite{art:skalaketal73,thesis:kruger}; $A_j$ is the surface are of the $j$-th element.
We adopt the Helfrich formulation to compute the free-energy $\WB$ related to the resistance to bending~\cite{art:helfrich73}. Following~\cite{thesis:kruger}, we discretise the {bending energy} as:
\begin{equation}\label{eq:helfrich}
\WB = \frac{\kB\sqrt{3}}{2}\sum_{\langle i,j\rangle}\left(\theta_{ij}-\theta_{ij}^{(0)}\right)^2\; ,
\end{equation}
where $\kB = 2\times 10^{-19}\mbox{ N m}$~\cite{book:gommperschick} is the bending modulus; the sum runs over all the neighbouring triangular elements, and $\theta_{ij}$ is the angle between the normals of the $i$-th and $j$-th elements ($\theta_{ij}^{(0)}$ is the same angle in the unperturbed configuration). Once we have the total free-energy $W = \WS+\WB$, we compute the force acting on the $i-$th node by performing the derivative of $W$ with respect to the coordinates of the node $\vec{x}_i$:
\begin{equation}\label{eq:nodal_force_energy}
\vec{F}_i = -\frac{\partial W(\vec{x}_i)}{\partial \vec{x}_i}\; .
\end{equation}
Note that we are implementing neither area nor volume conservation: in fact, as stated in \cite{D0SM00587H}, we checked that both area and volume were conserved, even without an explicit area or volume conservation law (see Electronic Supplementary Information in \cite{D0SM00587H}). \\
Regarding the viscoelastic term, we implement the Standard Linear Solid (SLS) model~\cite{art:lizhang19,li2020finite}. The viscous stress tensor is given by 
\begin{equation}\label{eq:mv2}
\pmb{\tau}^\nu =  \mus \left(2\dot{\vec{E}} -\mbox{tr}(\dot{\vec{E}})\mathbb{1}\right) + \mud \mbox{tr}(\dot{\vec{E}})\mathbb{1}\; ,
\end{equation}
where $\vec{E}$ is the strain tensor (see~\cite{art:lizhang19,D0SM00587H}); $\mus$ and $\mud$ are the shear and dilatational membrane viscosity: in this work, we assume $\mus~=~\mud~=~\mum$~\cite{D0SM00587H}. We refer to~\cite{art:lizhang19,D0SM00587H} for the computation of the stress tensor $\pmb{\tau}^\nu$ (Eq.~\eqref{eq:mv2}) as well as for the nodal force $\vec{F}_i$ (Eq.~\eqref{eq:nodal_force_energy}). \\
Finally, once we have the nodal force $\vec{F}_i$ for each node $i$ of the 3D mesh, we spread this force to the lattice nodes via the IBM (see~\cite{book:kruger} for details). We adopt the same scheme as in~\cite{D0SM00587H}.

\begin{figure*}[ht!]
    \centering
    \includegraphics[width=1.\linewidth]{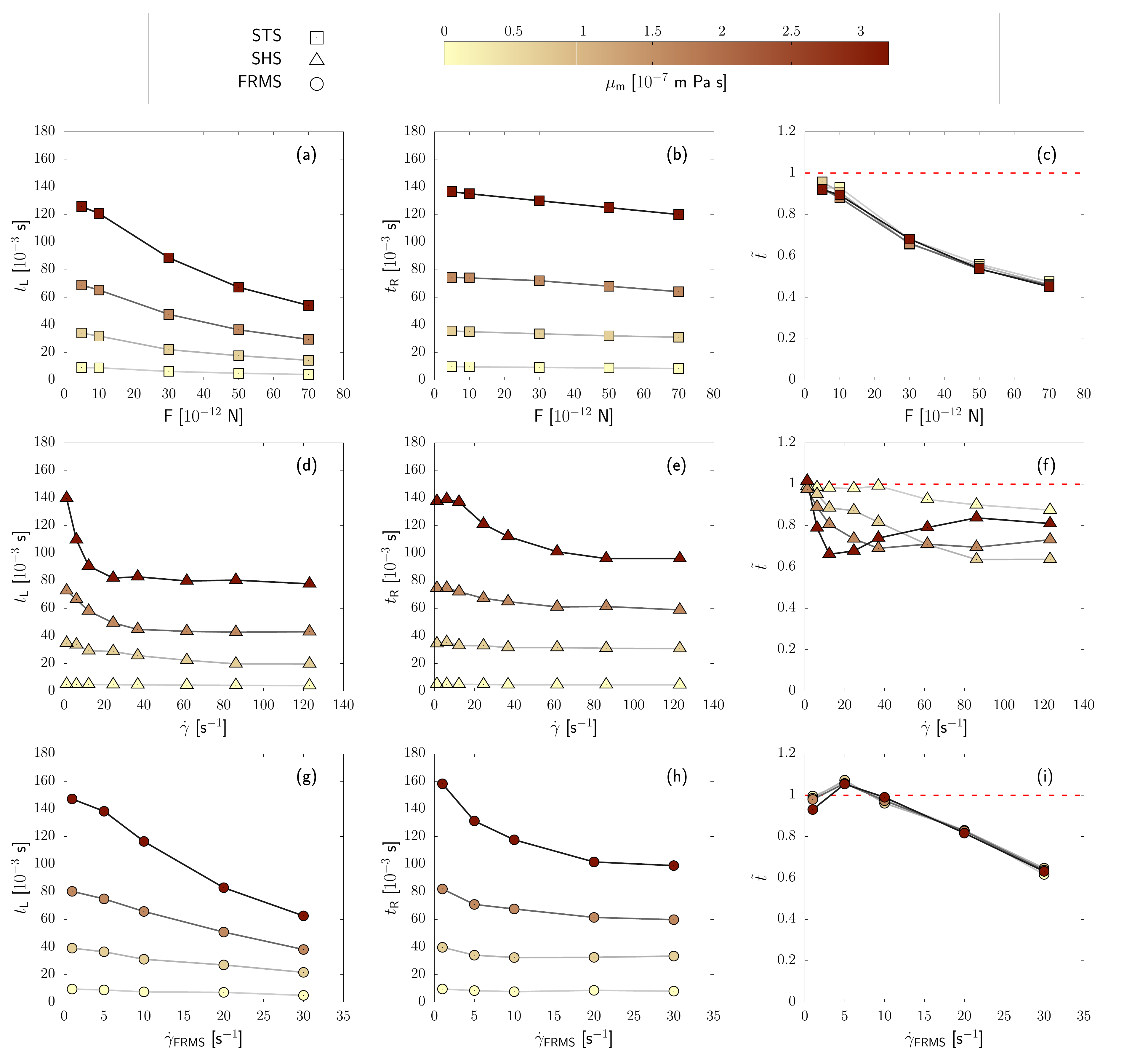}
    \caption{Characteristic times $\tl$ (first column of panels) and $\tr$ (second column of panels), as well as the ratio $\ttilde=\tl/\tr$ (third column of panels) are reported for the three simulations performed, i.e., stretching simulation (STS, \protect \Esquare, panels (a-c), Sec.~\ref{sec:stretching}), shear simulation (SHS, \protect \Etriangle, panels (d-f), Sec.~\ref{sec:shear}), four-roll mill simulation (FRMS, \protect \Ecircle, panels (g-i), Sec.~\ref{sec:four-roll_mill}), for different values of membrane viscosity $\mum$ (from lightest to darkest color): $\mum=0$~m~Pa~s (\protect \Lsquare, \protect \Ltriangle, \protect \Lcircle), $\mum=0.64 \times 10^{-7}$~m~Pa~s (\protect \LGsquare, \protect \LGtriangle, \protect \LGcircle), $\mum=1.59 \times 10^{-7}$~m~Pa~s (\protect \DGsquare, \protect \DGtriangle, \protect \DGcircle), $\mum=3.18 \times 10^{-7}$~m~Pa~s (\protect \Ssquare, \protect \Striangle, \protect \Scircle). The red dashed line represents the reference value for the symmetric case, i.e., $\ttilde=1$.\label{fig:t_vs_simulations}}
\end{figure*}
\section{Loading and relaxation time}\label{sec:loading_relaxation}	
In this section, we quantitatively study the loading time $\tl$ and the relaxation time $\tr$ with three different loading mechanisms (see Fig.~\ref{fig:sketch}): the stretching with optical tweezers (STS, see Sec.~\ref{sec:stretching}), the deformation in simple shear flow (SHS, see Sec.~\ref{sec:shear}) and the deformation in an elongational flow (FRMS, see Sec.~\ref{sec:four-roll_mill}). 
These three simulations differ mainly for two aspects (summarised in Tab.~\ref{tab:summary}): the first one is that the membrane can be deformed by an external force that acts directly on the membrane (like in the STS) or by the viscous friction with the fluid (SHS and FRMS); moreover, the membrane can rotate (like in the SHS) or not (STS and FRMS). 
The idea underlying the choice of these three different setups is to catch which of the aforementioned characteristics affects the loading and relaxation dynamics. \\
To quantify the loading time $\tl$ and the relaxation time $\tr$, we define the deformation parameter
\begin{equation}\label{eq:d}
D(t) = \frac{\dA(t)-\dT(t)}{\dA(t)+\dT(t)}\; ,
\end{equation}
where $\dA$ and $\dT$ represent the length of the axial and transversal diameters, i.e., the greater and medium eigenvalues of the inertia tensor (see~\cite{D0SM00587H}). 
In our computational domain, $\dA$ and $\dT$ lie in the $x-y$ plane. We define the average deformation $\dav$, i.e., the value of the deformation $D$ such that $\lim_{t\to \infty}D(t)=\dav$ in the loading simulation and $D(0)=\dav$ in the relaxation simulation.\\
Qualitatively, the loading time $\tl$ is the characteristic time the deformation $D(t)$ takes to reach $\dav$; the relaxation time $\tr$ is the characteristic time needed to relax to the initial shape, after the arrest of the mechanical load. 
Quantitatively, we can get $\tl$ and $\tr$ via a fit of $D(t)/\dav$ with the following functions:
\begin{equation}\label{eq:lst}
\lst(t) = 1 - \exp\left\{-\left(\frac{t}{\tl}\right)^{\deltal}\right\}\; ,
\end{equation}
\begin{equation}\label{eq:lsh}
\lsh(t) = 1 - \exp\left\{-\left(\frac{t}{\tl}\right)^{\deltal}\right\}\ \cos\left(\frac{t}{\tlcos}\right)\; ,
\end{equation}
\begin{equation}\label{eq:r}
R(t) = \exp\left\{-\left(\frac{t}{\tr}\right)^{\deltar}\right\}\; ,
\end{equation}
where $\lst$ is used to fit the loading time for the STS and the FRMS (see Sec.~\ref{sec:stretching}-\ref{sec:four-roll_mill}, respectively); $\lsh$ is used to fit the loading time for the SHS (see Sec.~\ref{sec:shear}); $R$ is used to fit the relaxation time $\tr$ for all three simulations; $\deltal$ and $\deltar$ are parameters introduced to improve the fit~\cite{fedosov2010multiscale} (see~\cite{D0SM00587H} for some more details) and will be characterised in Sec.~\ref{sec:comparison}.\\
Note that we propose two different functions to fit data during the loading (i.e., Eq.~\eqref{eq:lst} and Eq.~\eqref{eq:lsh}); this is due to the different deformation process of the RBC: in the STS and FRMS, $D(t)$ is a monotonic increasing function with an asymptote  in $D = \dav$ (see Fig.~\ref{fig:sketch}, panels (b) and (f)); in the SHS, $D(t)$ oscillates around $\dav$, and the amplitude of the oscillations varies with the value of the membrane viscosity $\mum$ (see Fig.~\ref{fig:sketch}, panel (d)). These oscillations have been also observed for viscoelastic capsules\cite{art:yazdanibagchi13,art:lizhang19,D0SM00587H,barthes1985}. Before starting the relaxation, for the STS and FRMS, we waited long enough to achieve the steady value of the deformation $\dav$;  for the SHS, we ensured the oscillations were small enough if compared to $\dav$. Notice, however, that depending on the importance of viscous effects with respect to elastic ones, there may be cases where such oscillations are damped on very long times~\cite{barthes1985,rallison1980note,art:yazdanibagchi13}.\\
Moreover, while the values of the external mechanical loads we simulated are comparable to each other in terms of stress (see Sec.~\ref{sec:comparison}), the values of $\dav$ for the SHS are much smaller than for the STS and FRMS (see Fig.~\ref{fig:d_and_phi}, panel (c)).\\
Since we want to compare the loading time $\tl$ and the relaxation time $\tr$, we define the ratio
\begin{equation}\label{eq:t_tilde}
\tilde{t} = \frac{\tl}{\tr}.
\end{equation}
In all the following simulations, the membrane viscosity ranges between $\mum\in[0,3.18]\times 10^{-7}\mbox{ m Pa s}$~\cite{evans1976membrane,chien1978theoretical,hochmuth1979red,tran1984determination,art:baskurt96,riquelme2000determination,art:tomaiuolo11,braunmuller2012hydrodynamic,art:prado15,fedosov2010multiscale}. In the STS, the applied force is in the range $F\in[5,70]\times 10^{-12}\mbox{ N}$; in the SHS, we simulated shear rates in the range $\gammadot\in[1.23, 123]\mbox{ s}^{-1}$; finally, in the FRMS, we simulated $\froll\in[1,120]\mbox{ s}^{-1}$ (see Eq.~\eqref{eq:force_FRMS}).
Values for all parameters used in the simulations in both physical and lattice units are reported in Tab.~1 in ESI\dag.\\

\subsection{Stretching simulation (STS)}\label{sec:stretching}
In order to simulate the stretching with optical tweezers, we apply two forces with the same intensity $F$ and opposite directions at the ends of the RBC (see Fig.~\ref{fig:sketch}, panel (a)).
Simulations are performed in a 3D box $\Lx\times\Ly\times\Lz=(28,12,12)\times 10^{-6}$~m.
In Fig.~\ref{fig:t_vs_simulations}, we report the loading time $\tl$ (panel (a)) and the relaxation time $\tr$ (panel (b)) as a function of $F$, for different values of membrane viscosity $\mum$. In both cases, increasing the loading strength (as well as decreasing the value of membrane viscosity $\mum$)  results in faster dynamics. It is interesting to compare $\tl$ and $\tr$: in Fig.~\ref{fig:t_vs_simulations}, panel (c), we report the ratio $\tilde{t}$ (see Eq.~\eqref{eq:t_tilde}). 

\subsection{Shear simulation (SHS)}\label{sec:shear}
In the shear simulation, the deformation is due to a linear shear flow with intensity $\dot{\gamma}$. We set the wall velocity $\Uw =  (\pm\dot{\gamma}\Lz/2,0,0)$ on the two plane walls at $z=\pm \Lz/2$, and the RBC is oriented as reported in Fig.~\ref{fig:sketch}, panel (c). 
This choice is surely the optimal one for the purpose of the present study, since we can focus on the relaxation/loading process without any further complication. In real-world experiments, indeed, RBCs do not necessarily orient in the  shear plane and evolve into a complex dynamics with many dynamical modes~\cite{minetti2019dynamics,cordasco2017shape}. In particular, for the values of shear rate $\gammadot$ we are interested in, a rolling dynamics is expected if the RBC is not oriented in the shear plane~\cite{mauer2018flow}. Further complications can be introduced by polydispersity, i.e., that RBCs may have different sizes and viscoelastic properties~\cite{art:suresh2005connections,dell1983molecular}. We have preferred to avoid all these extra complications which could not allow us for a fair assessment in the comparison between a pure relaxation dynamics and a pure loading dynamics.
Simulations are performed in a 3D box $\Lx\times\Ly\times\Lz=(20,32,20)\times 10^{-6}$~m.
In Fig.~\ref{fig:t_vs_simulations}, the loading time $\tl$ (panel (d)) and the relaxation time $\tr$ (panel (e)) as a function of $\dot{\gamma}$ for different values of membrane viscosity $\mum$ are reported, as well as the ratio $\tilde{t}$ (panel (f)). While the relaxation time $\tr$ shows a similar behaviour compared to the STS (see Fig.~\ref{fig:t_vs_simulations}, panel (b)), a few more words are needed regarding the loading time $\tl$. Unlike the STS, now we have two characteristic times, that are $\tl$ and $\tlcos$ (see Eq.~\eqref{eq:lsh}): $\tl$ measures the time the membrane takes to reach the average deformation $\dav$, while $\tlcos$ measures the period of the oscillations. Data for $\tlcos$ are reported in ESI\dag, Fig.~1. In contrast to the STS, the loading time $\tl$ first decreases, and then it does not change much with the intensity of the mechanical load.

\begin{figure*}[ht!]
    \centering
    \includegraphics[width=1.\linewidth]{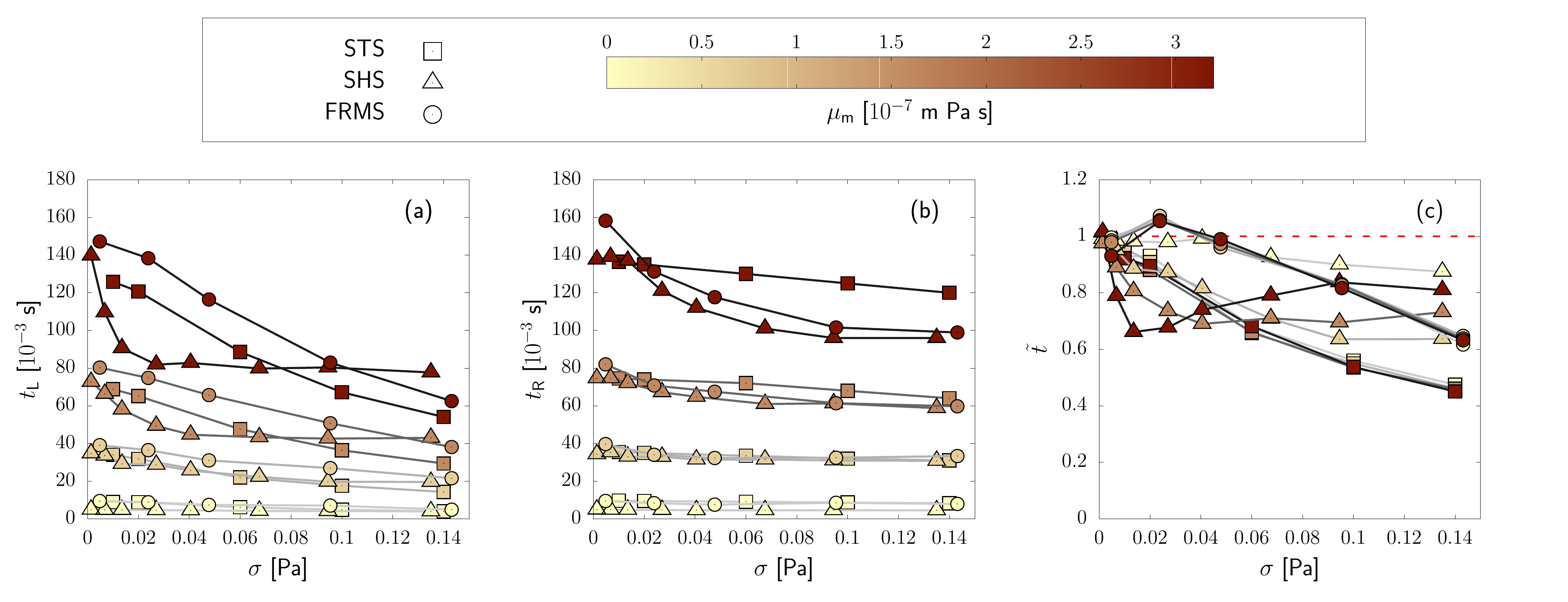}
    \caption{Comparison between the characteristic times $\tl$ (panel (a)), $\tr$ (panel (b)) and $\ttilde=\tl/\tr$ (panel (c)) as a function of the stress $\sigma$ (see Sec.~\ref{sec:comparison}) for the three simulations performed, i.e., stretching simulation (STS, \protect \Esquare, Sec.~\ref{sec:stretching}), shear simulation (SHS, \protect \Etriangle, Sec.~\ref{sec:shear}), four-roll mill simulation (FRMS, \protect \Ecircle, Sec.~\ref{sec:four-roll_mill}), for different values of membrane viscosity $\mum$ (from lightest to darkest color): $\mum=0$~m~Pa~s (\protect \Lsquare, \protect \Ltriangle, \protect \Lcircle), $\mum=0.64 \times 10^{-7}$~m~Pa~s (\protect \LGsquare, \protect \LGtriangle, \protect \LGcircle), $\mum=1.59 \times 10^{-7}$~m~Pa~s (\protect \DGsquare, \protect \DGtriangle, \protect \DGcircle), $\mum=3.18 \times 10^{-7}$~m~Pa~s (\protect \Ssquare, \protect \Striangle, \protect \Scircle). The red dashed line represents the reference value for the symmetric case, i.e., $\ttilde=1$.\label{fig:t_vs_sigma}}
\end{figure*}

\begin{figure}[ht!]
    \centering
    \includegraphics[width=.8\linewidth]{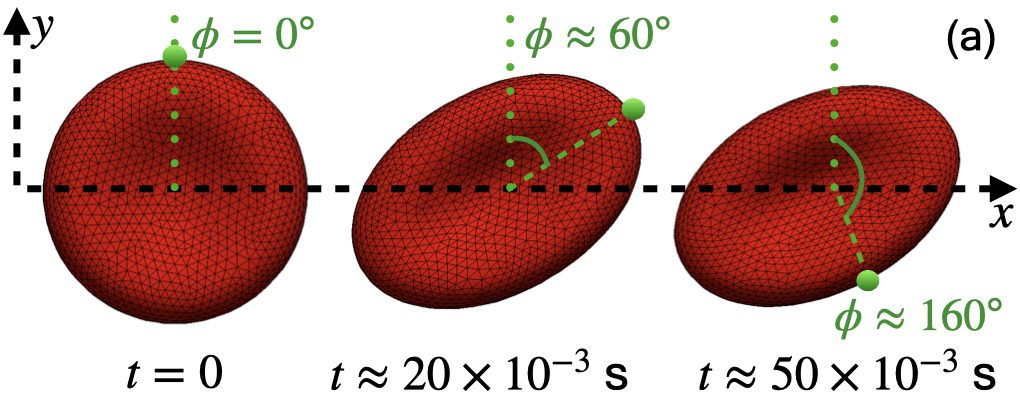}\vspace{.5cm}
    \includegraphics[width=.8\linewidth]{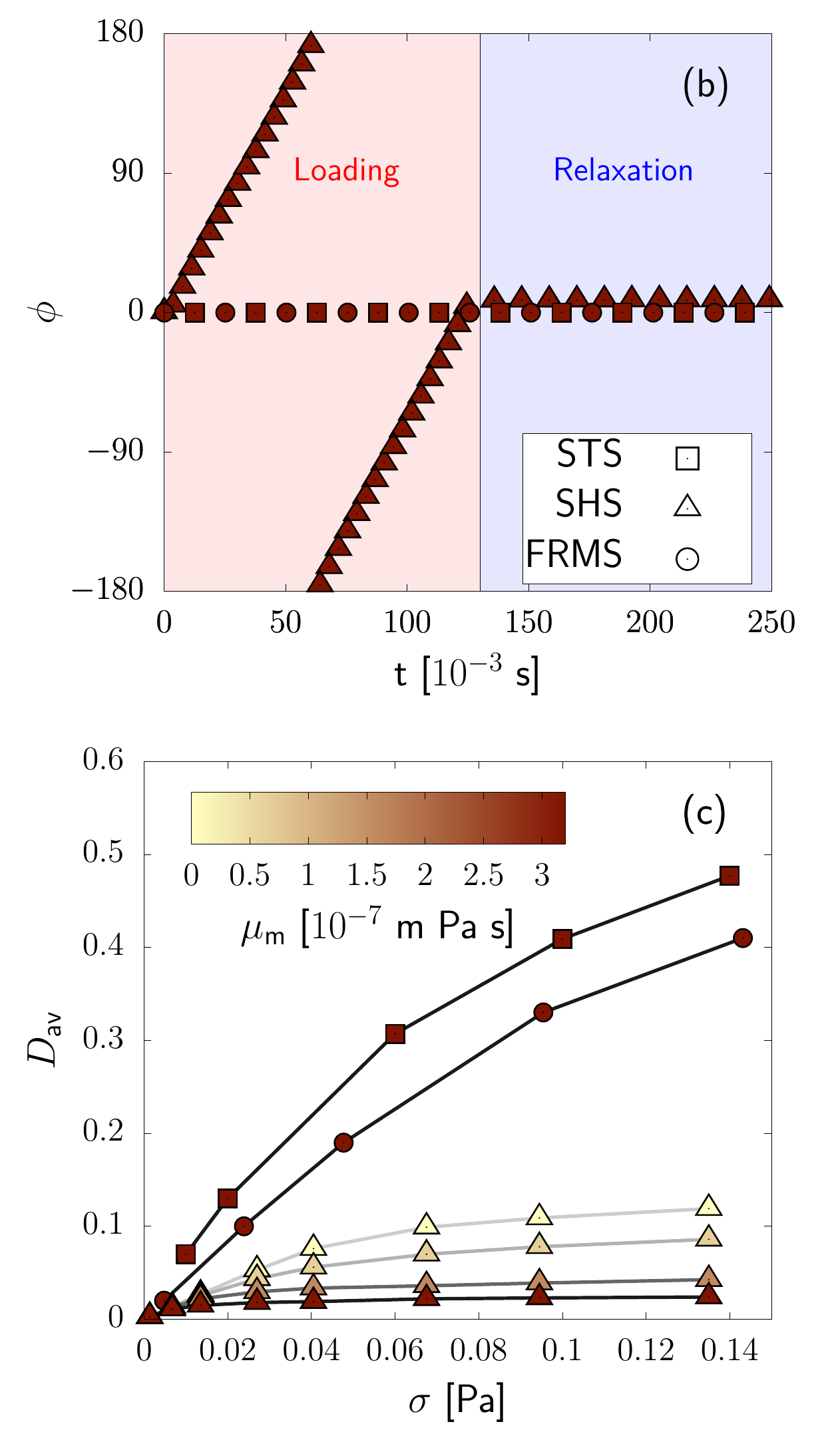}
    \caption{{\it\underline{Panel (a)}}: 
    snapshots of RBC at selected times. A point is selected on the membrane (green sphere) to perform a Lagrangian tracking and determine the time dependency of the angle $\phi$ that the point direction forms with the y axis in the deformation plane. {\it \underline{Panel (b)}}: we report the angle $\phi$ (see panel (a) above) as a function of time for the stretching simulation (STS, \protect \Ssquare, Sec.~\ref{sec:stretching}), shear simulation (SHS, \protect \Striangle, Sec.~\ref{sec:shear}), four-roll mill simulation (FRMS, \protect \Scircle, Sec.~\ref{sec:four-roll_mill}), for $\mum=3.18 \times 10^{-7}$~m~Pa~s. The red and blue shades represent loading and relaxation regions, respectively. {\it \underline{Panel (c)}}: average deformation $\dav$ (see text for details) as a function of the stress $\sigma$ for the three simulations performed (STS, SHS and FRMS). SHS data are displayed for different values of membrane viscosity $\mum$ (from lightest to darkest color):  $\mum=0$~m~Pa~s (\protect \Ltriangle), $\mum=0.64 \times 10^{-7}$~m~Pa~s (\protect \LGtriangle), $\mum=1.59 \times 10^{-7}$~m~Pa~s (\protect \DGtriangle), $\mum=3.18 \times 10^{-7}$~m~Pa~s (\protect \Striangle). STS data (\protect \Ssquare) and FRMS data (\protect \Scircle) data are only reported for $\mum=3.18 \times 10^{-7}$~m~Pa~s.
\label{fig:d_and_phi}}
\end{figure}

\subsection{Four-roll mill simulation (FRMS)}\label{sec:four-roll_mill}
In this case, we simulate the effect of four cylinders rotating~\cite{malaspinas2010lattice}, as shown in Fig.~\ref{fig:sketch}, panel (e), in order to create a flow similar to a pure elongational one. 
Simulations are performed in a 3D box $\Lx\times\Ly\times\Lz=(48,48,20)\times 10^{-6}$~m.
The idea is to simulate a loading mechanism that is a mixture of stretching with optical tweezers and deformation in simple shear flow (see Tab.~\ref{tab:summary}): in fact, in this case, the membrane does not rotate (like in the STS) and the deformation is caused by the flow (like in the SHS).\\
To create such a flow, we impose a force density~\cite{malaspinas2010lattice}
\begin{equation}\label{eq:force_FRMS}
\vec{F}(x,y) = 2k \mu \froll \left(\begin{array}{c}
\sin(kx)\cos(ky) \\
-\cos(kx)\sin(ky) \\
0
\end{array}\right)\; ,
\end{equation}
where $k=2\pi/\Lx$, $\mu$ is the local fluid viscosity, and $\froll$ is used to tune the load strength.  
We multiplied Eq.~\eqref{eq:force_FRMS} by $k$ to make the velocity gradient independent of the size of the fluid domain\footnote{The following result is valid in a homogeneous fluid with dynamics viscosity $\mu$.}:
\begin{equation}\label{eq:vel_grad_FRMS}
\frac{\partial \vec{u}}{\partial\vec{x}} = 
\froll\left(\begin{matrix}
\cos(kx)\cos(ky) &  -\sin(kx)\sin(ky) \\
\sin(kx)\sin(ky) & -\cos(kx)\cos(ky) 
\end{matrix}\right)\; ,
\end{equation}
where we have reported only $x$ and $y$ components, i.e., the components in the plane of the shear. Note that Eq.~\eqref{eq:force_FRMS} gives a pure elongational flow only in $x=\pi/2, 3\pi/2$ and in $y=\pi/2, 3\pi/2$.\\
In Fig.~\ref{fig:t_vs_simulations} we report the loading time $\tl$ (panel (g)) and the relaxation time $\tr$  (panel (h)) as a function of $\froll$. As for the STS and the SHS, both $\tl$ and $\tr$ decrease when the loading force increases or the membrane viscosity $\mum$ decreases. In Fig.~\ref{fig:t_vs_simulations}, panel (i), the ratio $\ttilde$ is reported. 

\section{Discussion\label{sec:comparison}}
In our simulations, the intensity of the three kinds of mechanical loads is changed by varying different quantities, i.e., $F$ for the STS, $\dot{\gamma}$ for the SHS and $\froll$ for the FRMS. To facilitate a comparison between them, we first consider the characteristic times $\tl$ and $\tr$ as well as the ratio $\ttilde$ as a function of the characteristic simulation stress $\sigma$ (Fig.~\ref{fig:t_vs_sigma}). To evaluate the stress $\sigma$ for the STS, we computed the area $A$ at the end of the RBC where the force $F$ is applied. Then, the stress is given by $\sigma^{\mbox{\scriptsize STS}}=F/A$; for the SHS, we wrote the stress as $\sigma^{\mbox{\scriptsize SHS}}=2\gammadot\muout$~\cite{thesis:kruger}. Finally, for the FRMS, the stress is given by the stress-peak $\sigma^{\mbox{\scriptsize FRMS}}=\muout\froll$. In all three simulations, the loading and relaxation times ($\tl$ and $\tr$, respectively) show qualitatively the same behaviour, i.e., they decrease when the loading strength increases or when the membrane viscosity $\mum$ decreases (see Fig.~\ref{fig:t_vs_sigma}, panels (a) and (b)); the ratio $\ttilde=\tl / \tr$ is reported in Fig.~\ref{fig:t_vs_sigma}, panel (c). For small forces ($\sigma\to 0$) we observe a clear tendency towards \textit{symmetry} between loading and relaxation ($\ttilde(\sigma,\mum) \rightarrow 1$), meaning that the characteristic times $\tl$ and $\tr$ tend to be equal. This is the limit where one expects to recover the {\it intrinsic} dynamics of the membrane, which depends only on the value of membrane viscosity $\mum$~\cite{D0SM00587H,art:prado15}. \\
On the other hand, for force strengths large enough, loading and relaxation dynamics are \textit{asymmetrical}, i.e., $\ttilde \neq 1$. As already noticed elsewhere~\cite{barthes2016motion}, this asymmetry could be explained by energetic considerations: in fact, during the loading phase, the deformation is driven by the external load (i.e., an external source of energy), while during the relaxation, the membrane provides the whole energy. Beyond these qualitative considerations, results in Fig.~\ref{fig:t_vs_sigma} provide a systematic characterisation of the relaxation times, as a function of either the stress $\sigma$ or the membrane viscosity $\mum$: an important message conveyed by our analysis is that the {\it asymmetry is not universal}, i.e., on equal values of membrane viscosity $\mum$, the ratio $\ttilde$ depends on the kind of mechanical load. Just to give some numbers, the difference between the values of $\ttilde$ for the STS and FRMS is roughly constant ($\approx$ 30\%) and it goes to zero for small values of $\sigma$; for the SHS the situation is a bit more complex because $\ttilde$ depends on $\mum$. However, if we compare SHS against STS for $\mum=3.18\times 10^{-7}$~m~Pa~s, we find a difference of less then 30\% for small values of $\sigma$ (i.e., $\sigma<0.1$~Pa), while
such a difference goes over the 50\% for large values of $\sigma$ (i.e., $\sigma>0.1$~Pa).\\ 
If we think that the asymmetry comes from the presence of a mechanical load with load strength large enough~\cite{diaz2000transient,barthes2016motion}, it comes natural to expect a non-universality and a dependency on the details of the loading mechanism. Thanks to our analysis, we are in a condition to further characterise this non-universality: indeed, we observe that while $\tilde{t}$ does not depend on $\mum$ for the STS and FRMS ($\ttilde^{\mbox{\scriptsize{STS}}} = \ttilde^{\mbox{\scriptsize{STS}}}(\sigma)$, $\ttilde^{\mbox{\scriptsize{FRMS}}} = \ttilde^{\mbox{\scriptsize{FRMS}}}(\sigma)$), it actually does in the SHS ($\ttilde^{\mbox{\scriptsize{SHS}}} = \ttilde^{\mbox{\scriptsize{SHS}}}(\sigma,\mum)$). The collapse shown by $\ttilde$ in the STS and FRMS (see Fig.~\ref{fig:t_vs_sigma}, panel (c)) suggests a factorisation of the loading and relaxation times in two contributions: one depending on the membrane viscosity $\mum$ and one on the load intensity $\sigma$: 
\begin{equation}\label{eq:tl_k}
\tl^K(\sigma,\mum)\approx\tl^{*K}(\sigma)t_0^K(\mum)\qquad\mbox{for }K=\mbox{STS, FRMS}\; ,
\end{equation}
\begin{equation}\label{eq:tr_k}
\tr^K(\sigma,\mum)\approx \tr^{*K}(\sigma)t_0^K(\mum)\qquad\mbox{for }K=\mbox{STS, FRMS}\; ,
\end{equation}
where the superscript $K$ stands for the kind of mechanical load. Given this factorisation, we have 
\begin{equation}\label{eq:ttilde_k}
\ttilde^K(\sigma) = \frac{\tl^{*K}(\sigma)}{\tr^{*K}(\sigma)}\qquad\mbox{for }K=\mbox{STS, FRMS}\; .
\end{equation}

\begin{figure*}[t!]
    \centering
    \includegraphics[width=1.\linewidth]{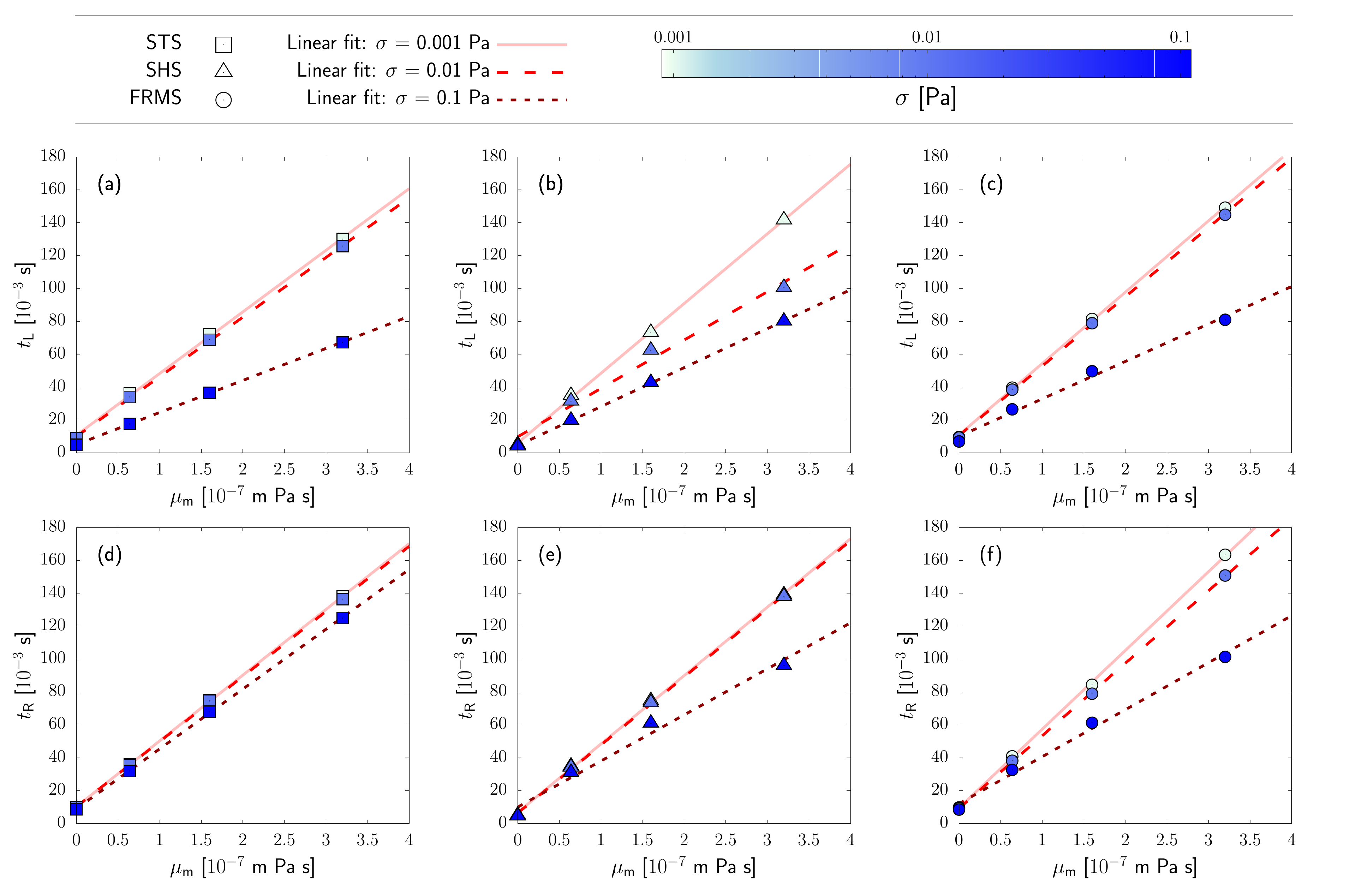}
    \caption{Characteristic times $\tl$ (panels (a-c)) and $\tr$ (panels (d-f)) as a function of the membrane viscosity $\mum$ (see Sec.~\ref{sec:comparison}) for the three performed simulations: stretching simulation (STS, \protect \Esquare, Sec.~\ref{sec:stretching}), shear simulation (SHS, \protect \Etriangle, Sec.~\ref{sec:shear}), four-roll mill simulation (FRMS, \protect \Ecircle, Sec.~\ref{sec:four-roll_mill}), for different values of stress $\sigma$ (from lightest to darkest color):
$\sigma=0.001$~Pa (\protect \LBsquare, \protect \LBtriangle, \protect \LBcircle),
$\sigma=0.01$~Pa (\protect \Bsquare, \protect \Btriangle, \protect \Bcircle),
$\sigma=0.1$~Pa (\protect \DBsquare, \protect \DBtriangle, \protect \DBcircle).\label{fig:t_vs_mu}}
\end{figure*}

\begin{figure*}[ht!]
    \centering
    \includegraphics[width=1.\linewidth]{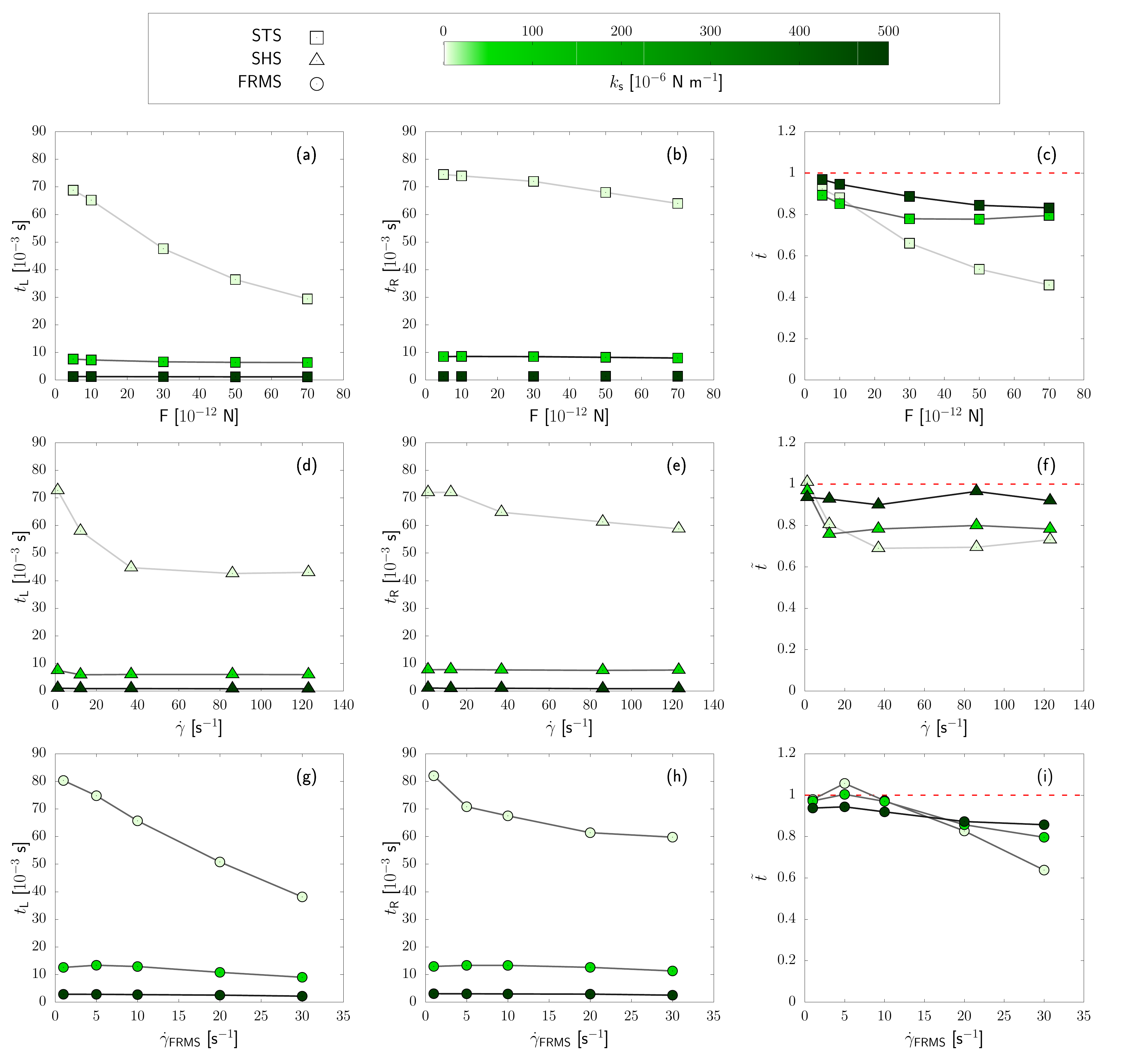}
    \caption{Characteristic times $\tl$ (first column of panels) and $\tr$ (second column of panels) as well as the ratio $\ttilde=\tl/\tr$ (third column of panels) are reported for the three simulations performed, i.e., stretching simulation (STS, \protect \Esquare, panels (a-c), Sec.~\ref{sec:stretching}), shear simulation (SHS, \protect \Etriangle, panels (d-f), Sec.~\ref{sec:shear}), four-roll mill simulation (FRMS, \protect \Ecircle, panels (g-i), Sec.~\ref{sec:four-roll_mill}), for different values of surface elastic shear modulus $\kS$ (from lightest to darkest color): $\kS=5.3\times 10^{-6}\mbox{ N m}^{-1}$ (\protect \LGRsquare, \protect \LGRtriangle, \protect \LGRcircle), $\kS=53\times 10^{-6}\mbox{ N m}^{-1}$ (\protect \GRsquare, \protect \GRtriangle, \protect \GRcircle), $\kS=530\times 10^{-6}\mbox{ N m}^{-1}$ (\protect \DGRsquare, \protect \DGRtriangle, \protect \DGRcircle). The value of membrane viscosity $\mum$ was kept fixed in all the simulations ($\mum=1.59\times 10^{-7}\mbox{ m Pa s}$).
    The red dashed line represents the reference value for the symmetric case, i.e., $\ttilde=1$.\label{fig:t_vs_simulations_ks}}
\end{figure*}

To make progress, we investigated what is the physical ingredient at the core of the factorisation given in Eqs.~\eqref{eq:tl_k}-\eqref{eq:tr_k}, or, alternatively, why the SHS does not factorize as in Eqs.~\eqref{eq:tl_k}-\eqref{eq:tr_k}. 
We investigated if such non-factorisation could be related to the oscillations of $D(t)$ that appear only in the SHS (see Sec.~\ref{sec:loading_relaxation}): however, since $\ttilde^{\mbox{\scriptsize{SHS}}}$ does not factorise even when the deformation $D(t)$ does not oscillate (e.g., for small values of the membrane viscosity $\mum$ and/or the shear rate $\gammadot$), we conclude that these oscillations cannot be fully responsible for the non-factorisation. We rather think that the difference between SHS and STS/FRMS is mainly due to the different dynamics that are induced by the mechanical load (see Tab.~\ref{tab:summary}).
In fact, regarding the SHS, one can split the velocity gradient $\frac{\partial \vec{u}}{\partial\vec{x}}$ in the symmetric (rotational) and antisymmetric (elongational) parts: 
\begin{equation}
\frac{\partial \vec{u}}{\partial\vec{x}} = 
\left(\begin{matrix}
0 & \dot{\gamma} \\
0 & 0
\end{matrix}\right)=
\left(\begin{matrix}
0 & \frac{\dot{\gamma}}{2} \\
\frac{\dot{\gamma}}{2} & 0
\end{matrix}\right) + \left(\begin{matrix}
0 & \frac{\dot{\gamma}}{2}  \\
-\frac{\dot{\gamma}}{2}  & 0
\end{matrix}\right)\; ,
\end{equation}
where the only two components in the shear plane are reported. 
The rotational part causes the rolling motion of the membrane (see Fig.~\ref{fig:d_and_phi}, panels (a) and (b)), while the elongational one tends to deform the RBC and pushes the main diameter to an angle of $\pi/4$ with respect to the shear direction; an increase in the membrane viscosity causes an increase in the time needed for the membrane to adapt to the flow and to deform; meanwhile, the rotational component promotes a rotation of the main diameter. Overall, the increase in membrane viscosity $\mum$ leads to a decrease of the average deformation $\dav$ (see also Fig.~\ref{fig:d_and_phi}, panel (c)). To make these arguments clearer, we have made two videos available in the ESI\dag: in one we show the simulation with $\gammadot=123$ s$^{-1}$ and $\mum = 3.18\times 10^{-7}\mbox{ m Pa s}$, while in the other one the simulation with $\gammadot=123$ s$^{-1}$ and $\mum = 0\mbox{ m Pa s}$ is reported. In both cases, the tank-treading motion of the membrane appears, but, for $\mum = 0\mbox{ m Pa s}$, the membrane deforms more than in the case with $\mum = 3.18\times 10^{-7}\mbox{ m Pa s}$. In Fig.~\ref{fig:d_and_phi}, panel (c), we report the average deformation $\dav$ as a function of the stress $\sigma$ for all three mechanical loads at changing the membrane viscosity $\mum$. As already observed in~\cite{D0SM00587H} for the STS, we found that $\dav$ is not sensitive to the value of membrane viscosity $\mum$. Moreover, in the FRMS we found that $\dav$ shows very little dependency on $\mum$, at least in the range of $\mum$ and $\sigma$ analysed. Hence, for STS and FRMS, we report only points for $\mum=3.18 \times 10^{-7}$~m~Pa~s. It emerges that, in the SHS, the average deformation $\dav$ saturates at a constant value: an increase in the shear rate $\gammadot$ causes an initial increase of the average deformation $\dav$; then, $\dav$ reaches a plateau and increasing the shear rate $\gammadot$ beyond a certain value does not result in an increased deformation. The higher the membrane viscosity $\mum$, the lower is the value of $\gammadot$ for which the plateau is reached. Furthermore, when compared to the STS and FRMS, we can see that on the same values of $\sigma$ the average deformation $\dav$ is much smaller in the SHS. Again, this is due to the rotation of the membrane during the loading. In both STS and FRMS, the membrane does not rotate, so that the energy injected by the flow is used to deform the membrane. These investigations reveal that it is impossible to predict the loading and relaxation times if we only know the deformation and have no information about the kind of mechanical load.\\
In view of the above considerations on the deformation, it appears also natural to study the characteristic times as a function of the average deformation $\dav$. We performed this analysis (see ESI\dag, Fig.~2), confirming the picture displayed in Fig.~\ref{fig:t_vs_sigma}: again, $\ttilde$ shows a collapse for the STS and the FRMS and does not depend on the value of the membrane viscosity $\mum$; for the SHS $\ttilde$ shows a dependency on both $\dav$ and the membrane viscosity $\mum$. The results on $\tl(\dav)$ and $\tr(\dav)$ (Fig.~2 in ESI\dag, panels (a) and (b), respectively) further confirm that in general there is no correlation between the degree of deformation of the membrane and the characteristic times for different kinds of mechanical loads.\\
In our previous work~\cite{D0SM00587H}, we have already seen that, for small forces, $\tr$ is linear in $\mum$, in agreement with literature predictions~\cite{art:prado15}. Now we can go further, and we study the dependency of both $\tl$ and $\tr$ as a function of $\mum$ for different values of $\sigma$. This will help further to determine to what degree these two kinds of dynamics can be regarded as different dynamics~\cite{diaz2000transient,barthes2016motion}. In Fig.~\ref{fig:t_vs_mu}, we report both $\tl(\mum)$ and $\tr(\mum)$ (first and second row of panels, respectively) for three values of $\sigma$ spanning two orders of magnitude as well as their linear fit (whose coefficients are reported in ESI\dag, Tab.~2) for all three simulations (STS in panel (a) and (d); SHS in panel (b) and (e); FRMS in panel (c) and (f)). In all three simulations, for a fixed value of $\sigma$, the linear approximation is reasonably good. For small values of $\sigma$ (e.g., $\sigma=0.001$~Pa), both $\tl$ and $\tr$ are similar for all three simulations; that is not surprising, since at small values of stress $\sigma$ the intrinsic properties of the membrane arise. Regarding the sensitivity of the linear trend with respect to a change in $\sigma$, we observe different behaviours in the two dynamics. Regarding the loading dynamics, we observe that for high values of $\sigma$, the three load mechanisms provide similar linear fits, while in the intermediate region of $\sigma$, the SHS shows a different behaviour than the STS and FRMS. Regarding the relaxation dynamics, the variability of the linear trends with the value of the stress is more pronounced in presence of hydrodynamical forces (i.e., SHS and FRMS), while in the STS the linear behaviour of $\tr$ with respect to $\mum$ is only slightly perturbed by a change in the stress if compared to the others. These quantitative observations are summarized in Tab.~2 in ESI\dag. \\
A dimensionless analysis could be performed to try to gain a deeper insight into the problem; however, if we try to make both the characteristic times $\tl$ and $\tr$ and the shear rates $\gammadot$ and $\froll$ dimensionless by using the characteristic elastic time $t_{\mbox{\tiny el}}=\muout r/\kS$, as well as the force $F$ by using the characteristic elastic force $F_{\mbox{\tiny el}}=r\kS$, we get only a rescaling along the x and y axis. Making the membrane viscosity $\mum$ dimensionless by introducing the Boussinesq number $\Bq=\mum/r\muout$ (see also~\cite{art:lizhang19,art:yazdanibagchi13,D0SM00587H}) results again in a constant rescaling of all the values, without giving a new insight into the problem.
In principle, one can look for some more refined non-dimensionalisation procedure combining membrane viscosity and rotational/deformation contributions: in the case of the SHS, for example, this would mean to find a shear time dependent on both $\Bq$ and $\Ca$. This would anyhow require a more precise knowledge (e.g. a phenomenological model~\cite{art:prado15}) on how the rotational/deformation contributions couple with the membrane viscosity effects.\\
Next, we discuss the parameters $\deltal$ and $\deltar$ used to improve the fit (see Eqs.~\eqref{eq:lst}-\eqref{eq:lsh}-\eqref{eq:r}). In all three setups, $\deltal$ and $\deltar$ are close to one, especially in the STS and FRMS (see Fig.~3 in ESI\dag). The biggest deviation can be found during the loading in the SHS, where the parameter $\deltal$ seems to tend asymptotically to $\deltal\approx 0.6$ at increasing values of shear rate $\gammadot$: this deviation from 1 reflects the effect of the kind of mechanical load also on $\deltal$, showing that, during the loading in the SHS, $D(t)$ is not that close to an exponential function, and then multiple time scales arise~\cite{D0SM00587H}. Indeed, having fitting parameters $\deltal$ and $\deltar$ different from 1 is symptomatic of the presence of multiple loading and relaxation times, respectively. This issue has already been investigated for both SHS and STS during the relaxation dynamics in our earlier study~\cite{D0SM00587H}; here we go deeper with the investigation during the loading dynamics. We have monitored the time evolution of the deformation $\frac{D(t)}{\dav}$ in log-lin scale (see Figs.~4-5-6 in ESI\dag). The initial stage of the loading/relaxation process is well characterised by a single "dominant" time scale, and only later, when the difference of $D(t)$ from $\dav$ (during the loading) or from 0 (during the relaxation) is less (or even much less) than about $10\%$, other time scales appear. As already pointed out in~\cite{D0SM00587H}, it is difficult to make quantitative assessments on the "late" dynamics, because then the deformation is close to its steady value (for the loading) and/or to the rest value (for the relaxation), and in this situation, discretisation errors could have more influence. One could make a deeper analysis of these multiple relaxation times and fit data with two (or more) characteristic times, both during loading and relaxation; however, we notice that the values of $\tl$ and $\tr$ found by fitting with the stretched exponential (i.e., by using $\deltal,\deltar\neq 1$) are in good agreement with the "dominant" time scale (see Figs.~4-5-6 in ESI\dag). On the other hand, the small difference between the fitted characteristic times and the "dominant" characteristic times also explains why the ratio $\ttilde$ becomes greater than 1 for one case in the FRMS (see Fig.~\ref{fig:t_vs_simulations} panel (i) and Fig.~7 in ESI\dag).\\ 
Before closing this section, we also discuss the effect of the surface elastic shear modulus $\kS$ (see Eq.~\eqref{eq:skalak}) on the loading and relaxation times. As already stated in the introduction, some pathologies can affect the value of membrane elasticity~\cite{art:kruger14deformability, art:suresh2005connections,suresh2006mechanical,brandao2003optical,briole2021molecular,brandao2003elastic,fedosov2011quantifying,luo2013inertia,ye2013stretching,hosseini2012malaria}: for example, for RBCs infected with the malaria parasite Plasmodium falciparum, experiments with optical tweezers estimated values of elastic shear modulus ranging from $\kS = 5.3\times 10^{-6} \mbox{ N m}^{-1}$ (i.e., for the healthy RBC) to $\kS = 100\times 10^{-6} \mbox{ N m}^{-1}$~\cite{art:suresh2005connections}. For this purpose, we fixed the value of membrane viscosity $\mum=1.59\times 10^{-7}\mbox{ m Pa s}$ and we varied the elastic shear modulus $\kS$ in a range of two orders of magnitude: from $\kS = 5.3\times 10^{-6} \mbox{ N m}^{-1}$ to $\kS = 530\times 10^{-6} \mbox{ N m}^{-1}$. Results are reported in Fig.~\ref{fig:t_vs_simulations_ks}. For all three kinds of mechanical loads simulated, increasing the value of the elastic shear modulus by a factor 10 or 100 results in a reduction of the characteristic times by about the same factor, as expected~\cite{,evans1976membrane,art:prado15}. Moreover, we analysed the ratio $\ttilde$, finding that it gets closer to 1 at increasing values of $\kS$. Mechanical balance at the interface tells us that the relative importance of viscous to elastic effects rescales as the ratio between the shear rate $\gammadot$ developed in the fluid and the surface elastic modulus $\kS$ of the membrane. The shear rate developed in the fluid during both loading and relaxation is larger at increasing load strength; thus, for a fixed $\dot{\gamma}$, by increasing $\kS$ we should have the same results observed for $\kS = 5.3\times 10^{-6} \mbox{ N m}^{-1}$ at shear rates that are smaller by a factor given by the ratio of the $\kS$'s. In other words, we expect $\ttilde \rightarrow 1$ when $\kS$ gets very large. This fact is confirmed by our results. 
We hasten to remark that these are only preliminary results, for at least two reasons: first, a proper dimensionless analysis of the governing equations~\cite{art:yazdanibagchi13,luo2013inertia} reveals that also the importance of the bending modulus $\kB$ with respect to $\kS$ needs to be taken into account via a suitable dimensionless number $\kB^*$. 
Since it is not known whether certain blood-related pathologies, such as Plasmodium falciparum malaria parasite infection, affect the value of the bending modulus $\kB$~\cite{fedosov2011quantifying,luo2013inertia,ye2013stretching,hosseini2012malaria}, we have kept it fixed at the value for the healthy RBC (see Sec.~\ref{sec:method}). Therefore, in our simulations, at changing $\kS$, the dimensionless number $\kB^*=\frac{\kB}{r^2\ \kS}$ is changing: a more comprehensive analysis on the effects of $\kS$ should be done by assessing also the impact of a variation in $\kB^*$ separately.
Second, the SLS model we implemented to take into account the viscoelastic effects (see Sec.~\ref{sec:method}) contains an artificial elastic contribution $k'$ that is proportional to $\kS$ and needs to be tuned in order to recover physical results (see~\cite{D0SM00587H,art:lizhang19} for further details and validations). Varying the value of the elastic shear modulus $\kS$ modifies also the value of $k'$: if and how this affects the physical dynamics of the RBC at the very large $\kS$ we considered requires a detailed computational analysis by its own, which is out of the scope of the present paper. All these considerations surely warrant dedicated studies in the future.

\section{Conclusions\label{sec:conclusion}}
A comprehensive characterisation of the viscoelastic properties of the RBC membrane, as well as the way the membrane responds to an external force, is of paramount interest in different fields, from the detection of pathologies~\cite{art:kruger14deformability,art:suresh2005connections,art:prado15}, to the design of biomedical devices~\cite{murakami1979nonpulsatile,nonaka2001development,art:behbahani09,art:arora2006hemolysis}. A paradigmatic example is provided by ventricular assist devices~\cite{hassler2020finite} where RBCs evolve in a complex flow and their fate is closely linked to their residence time inside the device: if the residence time inside the impeller (that is the region where the RBCs experience a wide range of stress) is much shorter then the loading time, RBCs deform without reaching a steady state configuration; on the contrary, a higher residence time leads to a deformation that can cause hemolysis (that is, the release of the cytoplasm into surrounding plasma due to damage or rupture of the membrane).\\ 
In general, when an external force acts on a viscoelastic membrane, two main kinds of dynamics arise: the {\it loading} and the {\it relaxation} dynamics with associated times $\tl$ and $\tr$. Earlier investigations pointed to the fact that these two kinds of dynamics are two distinct processes, since during the relaxation there is no external force to drive the membrane, in contrast to the loading~\cite{diaz2000transient,barthes2016motion}. To the best of our knowledge, however, an exhaustive comparative characterisation of these two kinds of dynamics has never been conducted. This motivated our work to investigate these two kinds of dynamics with different setups that involve different typologies of mechanical loads (whose main features are summarised in Tab.~\ref{tab:summary}) while performing a parametric study on the values of membrane viscosity $\mum$. The latter choice is motivated by the large variability of membrane viscosity values reported in the literature~\cite{evans1976membrane,chien1978theoretical,hochmuth1979red,tran1984determination,art:baskurt96,riquelme2000determination,art:tomaiuolo11,braunmuller2012hydrodynamic,art:prado15,fedosov2010multiscale}.\\ 
The two kinds of dynamics are {\it symmetrical} ($\ttilde=\tl/\tr \rightarrow 1$) in the limit of small load strengths ($\sigma \rightarrow 0$), i.e., in the limit where the response function of the RBC is dominated by the "intrinsic" properties of the membrane; in marked contrast, we found an {\it asymmetry} in the two kinds of dynamics for load strengths large enough ($\ttilde=\tl/\tr\ne 1$ for $\sigma >0$), meaning that the loading dynamics is always faster than the relaxation one. We found that the asymmetry profoundly depends on the kind of mechanical load and we have demonstrated this non-universality via a quantitative study in terms of the applied load strength $\sigma$ and the value of membrane viscosity $\mum$. There are some realistic load mechanisms, like shear flows, that make the membrane rotate during loading while leaving the membrane relaxing to the shape at rest without rotation: in this case, the contribution that the membrane viscosity gives to the characteristic times $\tl$ and $\tr$ differs, and then the ratio $\ttilde$ is a function of both the stress $\sigma$ and the membrane viscosity $\mum$. 
From the other side, there are other realistic load mechanisms, like the stretching with optical tweezers or the deformation with an elongational flow, in which the membrane deforms without rotating during both processes. In this case, the contribution given by the membrane viscosity $\mum$ to the characteristic times is the same during both loading and relaxation, and as a consequence, the ratio $\ttilde$ is a function of the stress $\sigma$ only. Even though we showed that both loading and relaxation dynamics are not universal, we found that for a given value of the stress $\sigma$, a linear increase of the characteristic times as a function of the membrane viscosity $\mum$ is a fair approximation in all cases.\\
Finally, since some blood-related diseases~\cite{art:kruger14deformability, art:suresh2005connections,suresh2006mechanical,brandao2003optical,briole2021molecular,brandao2003elastic,fedosov2011quantifying,luo2013inertia,ye2013stretching,hosseini2012malaria} can alter the values of the elastic shear modulus $\kS$, we also investigated the loading and relaxation dynamics at changing $\kS$ for a fixed value of $\mum$, finding that larger values of $\kS$ promote symmetrization ($\ttilde \rightarrow 1$) of the dynamics.\\
We argue our findings offer interesting physical and practical insights on the response function and the unsteady dynamics of RBCs driven by realistic mechanical loads.

\section*{Conflicts of interest}
There are no conflicts to declare.

\section*{Acknowledgements}
The authors acknowledge L. Biferale and G. Koutsou. This project has received funding from the European Union Horizon 2020 Research and Innovation Program under the Marie Skłodowska-Curie grant agreement No. 765048. We also acknowledge support from the project ‘‘Detailed Simulation of Red blood Cell Dynamics accounting for membRane viscoElastic propertieS’’ (SorCeReS, CUP No. E84I19002470005) financed by the University of Rome ‘‘Tor Vergata’’ (‘‘Beyond Borders 2019’’ call).




\bibliography{rsc} 
\bibliographystyle{rsc} 

\end{document}